\documentclass[journal]{IEEEtran} % 双栏 journal
\usepackage[draft]{graphicx}
\usepackage{graphicx}
\usepackage{amsmath}
\usepackage{amssymb}
\usepackage{booktabs}
\usepackage{enumerate}
\usepackage[ruled]{algorithm2e}
\usepackage{algorithmic}
\usepackage{amsmath}
\usepackage{graphics}
\usepackage{epsfig}
\usepackage{subfigure}
\usepackage{mmap} %% 消除 fi问题
\usepackage{multirow}

\usepackage{array}

\usepackage{CJK}
\usepackage{mathtools}
\usepackage{bbding}
\usepackage{amsmath}
\usepackage{amssymb}
\usepackage{ntheorem}
\usepackage{graphicx}
\usepackage{xcolor}
\usepackage{stfloats}
\newcounter{TempEqCnt}

\psfull
\begin{document}
\title{{Robust Security Energy Efficiency Optimization for RIS-Aided  Cell-Free Networks with  Multiple Eavesdroppers}}%Physical Layer Security Transmission Based on
\author{Wanming Hao,~\IEEEmembership{Member,~IEEE,} Junjie Li, Gangcan Sun, Chongwen Huang,~\IEEEmembership{Member,~IEEE,} Ming Zeng,~\IEEEmembership{Member,~IEEE,} Octavia A. Dobre,~\IEEEmembership{Fellow,~IEEE},  Chau Yuen,~\IEEEmembership{Fellow,~IEEE}
\thanks{A Part of this paper has been submitted to 2023 IEEE ICC [1].}
\thanks{W. Hao,  J. Li and G. Sun are with the School of Electrical and Information Engineering, Zhengzhou University, Zhengzhou 450001, China.}
\thanks{C. Huang is with the College of Information Science and Electronic Engineering, Zhejiang University, Hangzhou 310027, China.}
\thanks{M. Zeng is with the Department of Electrical Engineering and Computer Engineering, Universite Laval, Quebec, G1V 0A6, Canada.}
\thanks{O. A. Dobre is with the Faculty of Engineering and Applied Science, Memorial University, St. Johns, NL A1B 3X5, Canada.}
\thanks{C. Yuan is with the Singapore University of Technology and Design, Singapore 487372, Singapore.}
}

\maketitle

\begin{abstract}
In this paper, we investigate the energy efficiency (EE) problem under reconfigurable intelligent surface (RIS)-aided secure cell-free networks, where multiple legitimate users and eavesdroppers (Eves) exist. We formulate a max-min secure EE optimization problem by jointly designing the distributed active beamforming and artificial noise at base stations as well as the passive beamforming at RISs under practical constraints.  To deal with it, we first divide the original optimization problem into two sub-ones, and then propose an iterative optimization algorithm to solve each sub-problem based on the fractional programming, constrained convex-convex procedure (CCCP) and semi-definite programming (SDP) techniques. After that, these two sub-problems are alternatively solved until convergence, and the final solutions are obtained. Next, we extend to the imperfect channel state information of the Eves' links, and investigate the robust SEE beamforming optimization problem by bringing the outage probability constraints. Based on this, we first transform the uncertain outage probability constraints into the certain ones by the bernstein-type inequality and sphere boundary techniques, and then propose an alternatively iterative algorithm to obtain the solutions of the original problem based on the S-procedure, successive convex approximation, CCCP and SDP techniques. Finally, the simulation results are conducted to show the effectiveness of the proposed schemes.
\end{abstract}

\begin{IEEEkeywords}
Cell-free networks, reconfigurable intelligent surface (RIS), security energy efficiency.
\end{IEEEkeywords}

\IEEEpeerreviewmaketitle
\setlength{\baselineskip}{1\baselineskip}
\newtheorem{definition}{Definition}
\newtheorem{fact}{Fact}
\newtheorem{assumption}{Assumption}
\newtheorem{theorem}{Theorem}
\newtheorem{lemma}{Lemma}
\newtheorem{corollary}{Corollary}
\newtheorem{proposition}{Proposition}
\newtheorem{example}{Example}
\newtheorem{remark}{Remark}

\section{Introduction}
With the rapid development of global mobile communications, the demand for high-speed traffic is increasing  [2]. To satisfy such huge requirement of wireless data, more base stations (BSs) should be deployed. However, this will lead to more serious interference among BSs [3]. To address this, a multi-point cooperative transmission-based  cell-free (CF) network structure can be developed, where users are served by multiple BSs simultaneously, effectively mitigating the interference [4]-[6]. Additionally, more BSs will also result in a large power consumption, which is undesirable for energy source [7]. Fortunately, the low-power and low-cost  reconfigurable intelligent surface (RIS) technique can be deployed nowadays, and it can smartly change the wireless circumstance so as to improve the signal's strength or wireless coverage via adjusting the reflecting coefficients (RCs) of RIS elements [8]-[10]. Deploying more RISs instead of BSs will improve the spectral efficiency (SE) and energy efficiency (EE) of the system in future.

Meanwhile, due to the characteristics of the broadcast and openness of wireless communications,  how to guarantee the information security is challenging [11], [12].  Recently, the physical layer security (PLS) is regarded as a promising technique to improve the wireless data security via designing beamforming and artificial  noise (AN) at the BS [13].
The decoding probability can be improved by reasonably aligning the beamforming and AN at the legitimate receiver. It can also conceal the illegal receiver's signal [14], [15].
It is worth noting that the recent application of the PLS technique to the CF networks and RIS-aided secure networks has aroused  great interest among researchers [16]-[18].

\subsection{State-of-the-Art}
Recently,  it has been verified that the CF networks can enhance the information security via cooperative beamforming optimization [19]-[22]. For example, to maximize the secrecy SE (SSE) of the full duplex CF networks, an effective double-loop algorithm was proposed [19]. [20] investigated the impact of hardware impairments on security performance in a CF network, and proposed a successive convex approximation (SCA) and path-following algorithm to maximize the SE of the system. [21] investigated the simultaneous wireless information and power transfer (SWIPT) in the secure CF networks, and derived the lower bound of ergodic SSE and average capture energy. [22] analyzed the security in a multiple-input multiple-output (MIMO) system based on the digital-to-analog converter (DAC) architecture, and derived the additive quantization noise model-based precise closed-form expression of the SE. After that, it proposed a path tracking-based power optimization algorithm to maximize the achievable SE.

Additionally, RIS is beneficial for the information security owning to its ability to enhance the signal strength of the legitimate users and weaken that of the eavesdroppers (Eves) by adjusting the RC of each RIS element [23]-[27].
For example, [23] studied the secure transmission problem in the multi-layer RIS-aided integrated terrestrial-aerial networks, and proposed a block coordinate reduction-based optimization framework.
[24] proposed for the first time to apply active RIS to PLS, and designed an alternative optimization (AO) algorithm, which effectively overcome the dual fading impact of the reflective link channel.
Using the stochastic geometry tool, [25] derived the exact closed-form expressions of probability density function and cumulative distribution function of the received signals  with/without RIS.
[26] considered a dual-function RIS, which can simultaneously transmit and reflect signals.  On this basis, the authors brought the non-orthogonal multiple access (NOMA) technique to the considered system and formulated a SSE maximization problem by jointly optimizing the active/passive beamforming and  split factors. Finally, an AO algorithm was developed by applying the SCA and semi-definite programming (SDP) techniques.
For enhancing the SSE of the aerial-ground communications, the RIS was brought in [27]. By jointly optimizing the trajectory and passive beamforming, the SSE of the system is effectively improved.

Based on the above analysis, the recent works mainly focus on the SE of the RIS-aided secure CF networks, and the SEE, as another important indicator, has not been investigated. Additionally, the above works all assume perfect channel state information (CSI). Although the legitimate users' CSI can be obtained by advanced channel estimation techniques, the Eves'  CSI is difficult to obtain for the BSs. Therefore, a more practical system model for the RIS-aided secure CF networks should be conducted when beamforming resources are studied.

\subsection{Main Contributions}
To fill in the above gap, in this paper we investigate the SEE problem in the RIS-aided secure CF networks based on perfect and imperfect CSI, and the main contributions are summarized below:

\begin{itemize}
 \item[$\bullet$]
 We consider a RIS-aided secure CF network, where multiple RISs and BSs are deployed. Meanwhile, we assume that there exist multiple Eves and legitimate users based on the practical scenario. To investigate the SEE, we formulate a max-min SEE optimization problem by jointly designing the cooperative active beamforming and AN at BSs as well as  the passive beamforming at RISs under the constraints of each BS transmit power and the unit modulo of each RIS element.
 \item[$\bullet$]
We first study the above problem based on perfect CSI to obtain the upper bound of the performance. In fact, it is extremely difficult to directly solve the formulated problem, and thus we divide it into two subproblems by fixing variables, namely the cooperative active beamforming and AN optimizing subproblem and the cooperative passive beamforming optimizing subproblem. For the former, the fractional programming (FP) approach is applied to transform the max-min fractional objective function into the subtractive form, and then we propose an iterative algorithm based on the SCA, constrained concave-convex procedure (CCCP) and SDP techniques to obtain the solutions. For the latter, a similar scheme is used to optimize the cooperative passive beamforming of the RISs. Finally, the two subproblems are updated alternatively until convergence, and the final solutions are obtained.
 \item[$\bullet$]
 Next, considering the difficulty to obtain perfect CSI for the eavesdropping links, we re-investigate the max-min SEE problem based on imperfect CSI in term of the Eves' links,  and introduce the outage probability constraint for the eavesdropping rate. To deal with it, we first apply the Bernstein-type inequality (BTI) and sphere boundary theories to transform the uncertain outage probability constraints into the certain ones. Next, we divide the transformed optimization problem into two subproblems, and propose an iterative algorithm to solve each one based on the S-procedure, SCA, CCCP and SDP techniques. Next, we obtain the solutions of the original problem by alternatively solving these two subproblems until convergence. Finally, we prove the convergence and analyze the computational complexity for the proposed algorithms.
\end{itemize}

The rest of this paper is organized as follows. In Section II, the system model of the downlink RIS-aided secure CF networks is introduced, and the max-min SEE optimization problem is formulated. Section III presents the proposed joint optimization framework based on the perfect CSI, and Section IV supplements the solutions based on imperfect CSI for the Eves' links. Section V analyzes the proposed algorithms. The simulation results are shown in Section VI. Finally, conclusions are given in Section VII.

\emph{Notation:}
Boldface letters denote vectors or matrices. $\mathbb{R}^{K\times J}$ and $\mathbb{C}^{K\times J}$ denote real-valued and complex-valued matrices of size $K\times J$, respectively. $\left[ {\cdot} \right]^T$, $\left[ {\cdot} \right]^H$, Tr$(\cdot)$, Rank($\cdot$), respectively, denote the transpose, conjugate transpose, trace and rank. For a complex value $x$, $\left| x \right|$ represents its modulus. For a vector $\mathbf{a}$, {diag}$\left( \mathbf{a} \right)$ represents the diagonal operation, $\left\| {\mathbf{a}} \right\|$ represents the Euclidean norm. For a matrix ${\bf{A}}$, ${\left[ {\mathbf{A}} \right]_{i,j}}$ is the element at position $\left( {i,j} \right)$,  ${\bf{A}}\succeq 0$ represents the semi-definite matrix, vec$\left( {\mathbf{A}} \right)$ represents the matrix vec operator of $\mathbf{A}$. $\mathbf{I}_M$ is an identity matrix of size $M\times M$. ${\left[ {\cdot} \right]^ + }$ denotes the $\max \left\{ {0,{\cdot}} \right\}$, and $\Pr \left\{ {\cdot} \right\}$ denotes the probability.

\begin{figure}[t]
  \centering
  \includegraphics[scale = 0.26]{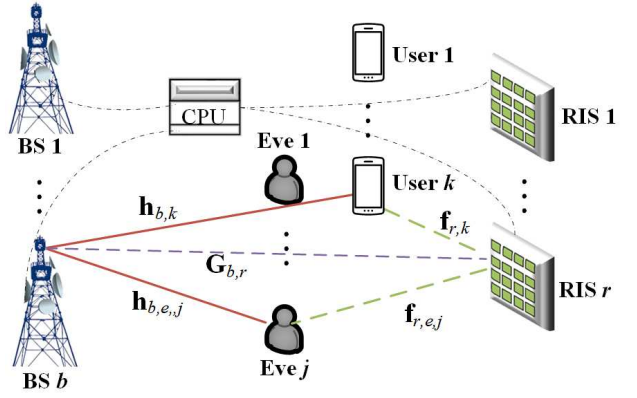}
  \caption{System model of the IRS-aided secure CF networks.}
  \label{fig.1}
\end{figure}

\section{System Model and Problem Formulation}
In this section, we first introduce the system model for the considered RIS-aided secure CF networks, and then formulate the max-min SEE optimization problem under practical constraints.
\subsection{System Model}
We consider a RIS-aided  downlink secure CF network, as shown in Fig. 1, which includes $B$ BSs and $R$ RISs. Here, we assume that there are $M$ antennas for each BS and $N$  reflective elements for each RIS for convenience. Additionally, we assume that there are $K$ single-antenna legitimate users and $J$ single-antenna Eves.
%We assume that there are $B$ BSs with $M$ antennas, and they are connected via high-speed backhaul links [22]. Meanwhile, there are $R$ RISs equipped with $N$ reflective elements.
%Additionally, there are $K$ legitimate users and $J$ Eves, and they are assumed as single-antenna devices.
Let $\forall b \in {\cal B} = \left\{ {1, \ldots ,B} \right\}$, $\forall r \in {\cal R} = \left\{ {1, \ldots ,R} \right\}$, $\forall n \in {\cal N} = \left\{ {1, \ldots ,N} \right\}$, $\forall k \in {\cal K} = \left\{ {1, \ldots ,K} \right\}$ and $\forall j \in {\cal J} = \left\{ {1, \ldots ,J} \right\}$ denote the sets of BSs, RISs,  RIS elements, users and Eves, respectively.

Each BS allocates a dedicated beamforming vector for each user, and thus the transmission signal of the $b$-th BS can be expressed as
\begin{equation}\label{eq:p1}
{{\bf{x}}_b} = \sum\nolimits_{k = 1}^K {\left( {{{\bf{w}}_{b,k}}{s_k} + {{\bf{v}}_{b,k}}} \right)},
\end{equation}
where ${{\bf{w}}_{b,k}} \in {\mathbb{C}^{M \times 1}}$ represents the beamforming vector of the $k$-th user at the $b$-th BS, ${s_k}$ represents the transmission symbol of the $k$-th user with $\left\{ {{{\left| {{s_k}} \right|}^2}} \right\} = 1$, and ${{\bf{v}}_{b,k}} \in {\mathbb{C}^{M \times 1}}$ represents the AN vector of the $b$-th BS for the $k$-th user [28].

Thus, the signals received by the $k$-th user and the $j$-th Eve wiretapping the $k$-th user's information are expressed as
\begin{subequations}\label{eq:p2}
\begin{align}
{y_k} &= \sum\limits_{b = 1}^B {\left( {{\bf{h}}_{b,k}^H + \sum\limits_{r = 1}^R {{\bf{f}}_{r,k}^H{\bf{\Theta }}_r^H{{\bf{G}}_{b,r}}} } \right)} {{\bf{x}}_b} + {n_k},\\
{y_{k,j}} &= \sum\limits_{b = 1}^B {\left( {{\bf{h}}_{b,e,j}^H + \sum\limits_{r = 1}^R {{\bf{f}}_{r,e,j}^H{\bf{\Theta }}_r^H{{\bf{G}}_{b,r}}} } \right)} {{\bf{x}}_b} + {n_{e,j}},
\end{align}
\end{subequations}
where ${\bf{h}}_{b,k}  \in {\mathbb{C}^{M \times 1}}$ and ${\bf{h}}_{b,e,j} \in {\mathbb{C}^{M \times 1}}$ denote the direct link channels from the $b$-th BS to the $k$-th user and the $j$-th Eve, respectively. ${\bf{G}}_{b,r}  \in {\mathbb{C}^{N \times M}}$, ${\bf{f}}_{r,k} \in {\mathbb{C}^{N \times 1}}$ and ${\bf{f}}_{r,e,j} \in {\mathbb{C}^{N \times 1}}$ denote the reflection link channels from $b$-th BS to the $r$-th RIS, the $r$-th RIS to the $k$-th user, and the  $r$-th RIS to the $j$-th Eve, respectively. ${n_k} \in {\cal C}{\cal N}\left( {0,\sigma _k^2} \right)$ and ${n_{e,j}} \in {\cal C}{\cal N}\left( {0,\sigma _{e,j}^2} \right)$ denote the complex additive white Gaussian noise of the $k$-th user and the $j$-th Eve, respectively, and own zero mean and variances ${\sigma _k^2}$ and ${\sigma _{e,j}^2}$, respectively. ${\bf{\Theta }}_r$ represents the phase shift matrix of the $r$-th RIS and can be written as ${{\mathbf{\Theta }}_r} =  {\text{diag}}\left( {{\theta _{r,1}}, \ldots ,{\theta _{r,N}}} \right)$, where ${\theta _{r,n}}$ is the $n$-th reflection element of the $r$-th RIS denoted as $\mathcal{F} = \left\{ {{\theta _{r,n}}\left| {{\theta _{r,n}} =  {\eta _{r,n}} {e^{j{\vartheta _{r,n}}}},{\vartheta _{r,n}} \in \left[ {0,2\pi } \right),{\eta _{r,n}} \in \left[ {0,1} \right]} \right.} \right\}$, and ${\eta _{r,n}}$ represents the reflection efficiency of the $n$-th reflection element of the $r$-th RIS. Without loss of generality, we set ${\eta _{r,n}} = 1$ to maximize the reflection signal [29].
Additionally, since the signals reflected via multiple times by RISs are very weak, we omit them here [30], [31].
% Here, we assume that the perfect CSI are available by advanced channel estimation techniques.

The desired signals for the $k$-th user and the $j$-th Eve who steals the information of the $k$-th user can be expressed as

\begin{subequations}\label{eq:p3}
\begin{align}
y_k^D &= \sum\limits_{b = 1}^B {\left( {{\mathbf{h}}_{b,k}^H + \sum\limits_{r = 1}^R {{\mathbf{f}}_{r,k}^H{\mathbf{\Theta }}_r^H{{\mathbf{G}}_{b,r}}} } \right)}{{\mathbf{w}}_{b,k}}{s_k}  \\
&= \left( {{\mathbf{h}}_{d,k}^H + {{\boldsymbol{\theta }}^H}{\mathbf{H}}_k^R} \right){{\mathbf{w}}_k}{s_k}  \\
&= {\hat{\boldsymbol \theta }}^H {{\mathbf{H}}_k}{{\mathbf{w}}_k}{s_k},  \\
y_{k,j}^D &= \sum\limits_{b = 1}^B {\left( {{\mathbf{h}}_{b,e,j}^H + \sum\limits_{r = 1}^R {{\mathbf{f}}_{r,e,j}^H{\mathbf{\Theta }}_r^H{{\mathbf{G}}_{b,r}}} } \right)} {{\mathbf{w}}_{b,k}}{s_k}  \\
&= \left( {{\mathbf{h}}_{d,e,j}^H + {{\boldsymbol{\theta }}^H}{{\bf{H}}_{e,j}^R}} \right){{\mathbf{w}}_k}{s_k}  \\
&= {\hat{\boldsymbol \theta }}^H {{\mathbf{H}}_{e,j}}{{\mathbf{w}}_k}{s_k},
\end{align}
\end{subequations}
where (3b) defines ${{\bf{h}}_{d,k}} = {\left[ {{\bf{h}}_{1,k}^T, \ldots ,{\bf{h}}_{B,k}^T} \right]^T}$, ${\boldsymbol{\theta }} = \left[ {{{\boldsymbol{\theta }}_1^T}, \ldots ,{{\boldsymbol{\theta }}_R^T}} \right]^T$, ${{\boldsymbol{\theta }}_r} = \text{diag}( {{{\bf{\Theta }}_r}} )$, ${\bf{H}}_k^R = \text{diag}\left( {{\bf{f}}_k^H} \right){\bf{G}}$, ${{\bf{f}}_k} = {\left[ {{\bf{f}}_{1,k}^T, \ldots ,{\bf{f}}_{R,k}^T} \right]^T}$, ${\bf{G}} = \left[ {{{\bf{G}}_1}, \ldots ,{{\bf{G}}_B}} \right]$, ${{\bf{G}}_b} = {\left[ {{\bf{G}}_{b,1}^T, \ldots ,{\bf{G}}_{b,R}^T} \right]^T}$, and ${{\bf{w}}_k} = {\left[ {{\bf{w}}_{1,k}^T, \ldots ,{\bf{w}}_{B,k}^T} \right]^T}$, (3c) defines ${\hat{\boldsymbol \theta }} = {\left[ {{{\boldsymbol{\theta }}^T},1} \right]^T}$ and ${{\bf{H}}_k} = {\left[ {{{( {{\bf{H}}_k^R} )}^H},{{\bf{h}}_{d,k}}} \right]^H}$, (3e) defines ${{\bf{h}}_{d,e,j}} = {\left[ {{\bf{h}}_{1,e,j}^T, \ldots ,{\bf{h}}_{B,e,j}^T} \right]^T}$, ${\bf{H}}_{e,j}^R = \text{diag}\left( {{\bf{f}}_{e,j}^H} \right){\bf{G}}$, and ${{\bf{f}}_{e,j}} = {\left[ {{\bf{f}}_{1,e,j}^T, \ldots ,{\bf{f}}_{R,e,j}^T} \right]^T}$, (3c) defines ${{\bf{H}}_{e,j}} = {\left[ {{{( {{\bf{H}}_{e,j}^R} )}^H},{{\bf{h}}_{d,e,j}}} \right]^H}$. For convenience, we define ${{\bf{w}}} = {\left[ {{\bf{w}}_{1}^T, \ldots ,{\bf{w}}_{k}^T} \right]^T}$,  ${{\bf{v}}} = {\left[ {{\bf{v}}_{1}^T, \ldots ,{\bf{v}}_{k}^T} \right]^T}$, and ${{\bf{v}}_k} = {\left[ {{\bf{v}}_{1,k}^T, \ldots ,{\bf{v}}_{B,k}^T} \right]^T}$.

On this basis,  the achievable secrecy rate of the $k$-th user can be expressed as
\begin{equation}\label{eq:p4}
{R_k} = {\left[ {{{\log }_2}\left( {1 + {\gamma _k}} \right) - \mathop {{\rm{max}}}\limits_{j \in {\cal J}} \left( {{{\log }_2}\left( {1 + {\gamma _{k,j}}} \right)} \right)} \right]^ + },
\end{equation}
where ${\gamma _k}$ and ${\gamma _{k,j}}$ denote the signal-to-interference-plus-noise ratio (SINR) of the $k$-th user and the $j$-th Eve wiretapping the information of the $k$-th user, respectively, which are given by
\begin{subequations}\label{eq:p5}
\begin{align}
{\gamma _k} &= \frac{{{{\left| {{{\hat{\boldsymbol \theta }}^H}{{\bf{H}}_k}{{\bf{w}}_k}} \right|}^2}}}{{\sum\limits_{i = 1,i \ne k}^K {{{\left| {{{\hat{\boldsymbol \theta }}^H}{{\bf{H}}_k}{{\bf{w}}_i}} \right|}^2} + \sum\limits_{i = 1}^K {{{\left| {{{\hat{\boldsymbol \theta }}^H}{{\bf{H}}_k}{{\bf{v}}_i}} \right|}^2} + } \sigma _k^2} }},\\
{\gamma _{k,j}} &= \frac{{{{\left| {{{\hat{\boldsymbol \theta }}^H}{{\bf{H}}_{e,j}}{{\bf{w}}_k}} \right|}^2}}}{{\sum\limits_{i = 1,i \ne k}^K {{{\left| {{{\hat{\boldsymbol \theta }}^H}{{\bf{H}}_{e,j}}{{\bf{w}}_i}} \right|}^2} + \sum\limits_{i = 1}^K {{{\left| {{{\hat{\boldsymbol \theta }}^H}{{\bf{H}}_{e,j}}{{\bf{v}}_i}} \right|}^2} + } \sigma _{e,j}^2} }}.
\end{align}
\end{subequations}

The total power consumption for the $k$-th user consists of transmit power and circuit power consumption, which can be calculated as
\begin{equation}\label{eq:p6}
P_k^{\text{total}} = \frac{1}{\zeta }\left( {{\rm{Tr}}\left( {{{\bf{w}}_k}{\bf{w}}_k^H} \right) \!+\! {\rm{Tr}}\left( {{{\bf{v}}_k}{\bf{v}}_k^H} \right)} \right) + {P_c},
\end{equation}
where ${\zeta }$ denotes the power amplifier efficiency, and ${P_c}$ denotes the circuit power consumption denoted as ${P_c} = B{P_B} + {P_U} + RN{P_R}$, where ${P_B}$, ${P_U}$ and ${P_R}$ are denoted as the hardware-dissipated power at the BS, user and RIS element, respectively. Finally, the SEE of the $k$-th user is defined as
\begin{equation}\label{eq:p7}
 \eta _k^{EE} = \frac{{{R_k}}}{P_k^{total}}.
\end{equation}

\subsection{Problem Formulation}
In this paper, we aim to maximize the minimum user's SEE by  jointly optimizing beamforming ${\bf{w}}$, AN ${\bf{v}}$ and phase shift ${\boldsymbol{\theta }}$ under the constraints of each BS transmit power and unit modulo at each RIS element, which can be formulated as
\begin{subequations}\label{eq:p8}
\begin{align}
 {{\cal P}_0}:& \mathop {\max }  \limits_{{\bf{w}},{\bf{v}},\boldsymbol{\theta} }  \  \mathop {\min }\limits_{k \in {\cal K}}  \  \eta _k^{EE}\\
 {\rm{s}}{\rm{.t.}}  \ & C1:   \sum\limits_{k = 1}^K {\left( {{\rm{Tr}}\left( {{{\bf{B}}_b}{{\bf{W}}_k}} \right) \!+\! {\rm{Tr}}\left( {{{\bf{B}}_b}{{\bf{V}}_k}} \right)} \right)}  \!\le\! {P_{b}}, \forall b \in {\cal B}, \\
& C2:  {\rm{       }}\left| {{\boldsymbol{\theta} _n}} \right| = 1,n = 1, \ldots ,RN,
\end{align}
\end{subequations}
where ${P_{b}}$ represents the maximum transmit power of the $b$-th BS, ${{\bf{W}}_k} = {{\bf{w}}_k}{{\bf{w}}_k}^H$, ${{\bf{V}}_k} = {{\bf{v}}_k}{{\bf{v}}_k}^H$, and ${{\bf{B}}_b} $ is defined as
\begin{equation}\label{eq:p9}
{{\bf{B}}_b} \buildrel \Delta \over = \text{diag}\left( {\underbrace {0, \ldots ,0}_{\left( {b - 1} \right)M},\underbrace {1, \ldots ,1}_M,\underbrace {0, \ldots ,0}_{\left( {B - b} \right)M}} \right), \forall b \in {\cal B}.
\end{equation}

Due to the non-convex constraints of the coupling of beamforming matrix ${\mathbf{W}}$, AN matrix ${\mathbf{V}}$ and phase shift vector $\boldsymbol{\theta}$ in ${{\cal P}_0}$, and the unit modulus constraint $C2$, it is difficult to solve ${\mathcal{P}_0}$.

\section{Proposed Joint Optimization Scheme}
To deal with ${{\cal P}_0}$, we propose an alternatively iterative scheme, namely solving $\boldsymbol{\theta}$ and $({\mathbf{W}},{\mathbf{V}})$ alternatively until convergence. We will give the detailed procedure next.
%\subsection{Overview of the proposed joint optimization scheme}
\subsection{Fix $\boldsymbol{\theta}$ and Solve $({\mathbf{W}},{\mathbf{V}})$}
%Here we propose a joint optimization scheme by alternately iterating ${\mathbf{W}}$, ${\mathbf{V}}$, $\boldsymbol{\theta}$ until converges.
We first introduce two auxiliary variables $\boldsymbol{\alpha} \in {\mathbb{R}^{K \times 1}}$ with $\boldsymbol{\alpha}  = \left[ {{\alpha _1}, \ldots ,{\alpha _K}} \right]$ and $\boldsymbol{\beta}  \in {\mathbb{R}^{K \times J}}$ with $\boldsymbol{\beta}  = \left[ {{\boldsymbol{\beta} _1}, \ldots ,{\boldsymbol{\beta} _K}} \right]$ , where ${\boldsymbol{\beta} _k} = \left[ {{\beta _{k}^1}, \ldots ,{\beta _{k}^J}} \right]$. Consequently, for given ${\boldsymbol{\theta}}^{[t]}$, ${\mathcal{P}_0}$ can be converted to the following one

\begin{subequations}\label{eq:p10}
\begin{align}
  {\mathcal{P}_1}&: \mathop {\max }\limits_{{\mathbf{W}},{\mathbf{V}},\boldsymbol{\alpha} ,\boldsymbol{\beta} } \mathop {\min }\limits_{k \in \mathcal{K}} \frac{{{{\log }_2}\left( {1 + {\alpha _k}} \right) - \mathop {{\text{max}}}\limits_{j \in \mathcal{J}} \left( {{{\log }_2}\left( {1 + {\beta _{k}^j}} \right)} \right)}}{{\frac{1}{\zeta }\left( {{\text{Tr}}\left( {{{\mathbf{W}}_k}} \right) + {\text{Tr}}\left( {{{\mathbf{V}}_k}} \right)} \right) + {P_c}}}  \\
  {\text{s}}{\text{.t}}{\text{.   }} & C3:  {\alpha _k} \leqslant \frac{{{{\hat {\boldsymbol \theta }}^{[t]H}}{{\mathbf{H}}_k}{{\mathbf{W}}_k}{\mathbf{H}}_k^H{\hat{\boldsymbol \theta }}^{[t]}}}{{{{{\hat{\boldsymbol \theta }}}^{[t]H}}{{\mathbf{H}}_k}{{\mathbf{L}}_k}{\mathbf{H}}_k^H{\hat{\boldsymbol \theta }}^{[t]} + \sigma _k^2}},\forall k \in \mathcal{K},  \\
  &  C4: \beta _k^j \geqslant \frac{{{{{\hat{\boldsymbol \theta }}}^{[t]H}}{{\mathbf{H}}_{e,j}}{{\mathbf{W}}_k}{\mathbf{H}}_{e,j}^H{\hat{\boldsymbol \theta }}^{[t]}}}{{{{{\hat{\boldsymbol \theta }}}^{[t]H}}{{\mathbf{H}}_{e,j}}{{\mathbf{L}}_k}{\mathbf{H}}_{e,j}^H{\hat{\boldsymbol \theta }}^{[t]} + \sigma _{e,j}^2}},\forall k \in \mathcal{K},\forall j \in \mathcal{J},  \\
  & C5:{\rm{Rank}}\left( {{{\bf{W}}_k}} \right) = 1,{{\bf{W}}_k}\succeq0,\forall k \in \mathcal{K},\\
  & C6: {\rm{Rank}}\left( {{{\bf{V}}_k}} \right) = 1,{{\bf{V}}_k}\succeq0,\forall k \in \mathcal{K},\\
  &  C1,
\end{align}
\end{subequations}
where ${{\mathbf{L}}_k} = \sum\nolimits_{i = 1,i \ne k}^K {{{\mathbf{W}}_k} + \sum\nolimits_{i = 1}^K {{{\mathbf{V}}_k}} }$. Since the objective function of ${\mathcal{P}_1}$ is not smooth enough, we introduce an auxiliary variable $z$ to reformulate ${\mathcal{P}_1}$ as
\begin{subequations}\label{eq:p11}
\begin{align}
  {\mathcal{P}_2} \ & : \mathop {\max }\limits_{{\mathbf{W}},{\mathbf{V}},\boldsymbol{\alpha} ,\boldsymbol{\beta},{z} } \ z \\
  {\text{s}}{\text{.t}}{\text{.}} \; C7 & : \frac{{{{\log }_2}\left( {1 + {\alpha _k}} \right) - {{\log }_2}\left( {1 + {\beta _{k}^j}} \right)}}{\frac{1}{\zeta }\left( {{\text{Tr}}\left( {{{\mathbf{W}}_k}} \right) + {\text{Tr}}\left( {{{\mathbf{V}}_k}} \right)} \right) + {P_c}} \geqslant z,\forall k \in \mathcal{K}, \nonumber \\
  & \qquad \qquad \qquad \qquad \qquad \qquad \qquad \quad \ \  \forall j \in \mathcal{J},\\
  C1 & ,   \ C3, \ C4,   \ C5, \ C6.
\end{align}
\end{subequations}
The molecule of $C7$ is a convex difference function, and the first-order Taylor approximation can be used to transform it into the following convex one
\begin{equation}\label{eq:p12}
\begin{split}
   &{\log _2}\left( {1 + {\alpha _k}} \right) - {\log _2}\left( {1 + \beta _k^j} \right)  \\
    {\approx} \ &  {\log _2}\left( {1 + {\alpha _k}} \right)  - {\log _2}\left( {1 + \beta {{_k^j}^{\left[ t \right]}}} \right) - \frac{{\beta _k^j - \beta {{_k^j}^{\left[ t \right]}}}}{{1 + \beta {{_k^j}^{\left[ t \right]}}}} \\
     \buildrel \Delta \over = \ &  f\left( {{\alpha _k},\beta _k^j,\beta {{_k^j}^{\left[ t \right]}}} \right).
\end{split}
\end{equation}
where $\left[ t \right]$ represents the $t$-th iteration. Thus, $C7$ can be rewritten as $\bar C7$, namely
\begin{equation}\label{eq:p13}
\bar C7:\frac{{f\left( {{\alpha _k},\beta _k^j,\beta {{_k^j}^{\left[ t \right]}}} \right)}}{{\frac{1}{\zeta }\left( {{\text{Tr}}\left( {{{\mathbf{W}}_k}} \right) \!+\! {\text{Tr}}\left( {{{\mathbf{V}}_k}} \right)} \right)\! +\! P_c}} \!\geqslant\! z,\forall k \in \mathcal{K},\forall j \in \mathcal{J}.
\end{equation}

Due to the non-convex constraints of the coupling of  ${\mathbf{W}}$ and ${\mathbf{V}}$  in $C3$, $C4$, and the fractional constraint of $\bar C7$, it is still difficult to solve ${\mathcal{P}_2}$.
Next, we use the FP method to convert the fractional constraint $\bar C7$ into a subtractive form equivalently [32]. It is obvious $f\left( {{\alpha _k},\beta _k^j,\beta {{_k^j}^{\left[ t \right]}}} \right) > 0$ and $\frac{1}{\zeta }\left( {{\text{Tr}}\left( {{{\mathbf{W}}_k}} \right) + {\text{Tr}}\left( {{{\mathbf{V}}_k}} \right)} \right) + {P_c} > 0$, then we introduce an auxiliary variable ${\boldsymbol{\rho }} \in {\mathbb{R}^{K \times J}}$ with ${\boldsymbol{\rho }} = \left[ {{{\boldsymbol{\rho }}_1}, \ldots ,{{\boldsymbol{\rho }}_K}} \right]$ and ${{\boldsymbol{\rho }}_k} = \left[ {{\rho }_k^1, \ldots ,{\rho }_k^J} \right]$, and $\bar C7$ can be rewritten as
\begin{equation}\label{eq:14}
\begin{split}
\tilde C7\!:\!f\left( {{\alpha _k}\!,\!\beta_k^j\!,\!\beta {{_k^j}^{\left[ t \right]}}\!,\!{\rho }_k^j} \right) &\!= \!2{{\rho }_k^j} \sqrt {f\left( {{\alpha _k},\beta _k^j,\beta {{_k^j}^{\left[ t \right]}}} \right)} \\
&\!-\! {{{\rho }_k^j}^2}  \left( {\frac{1}{\zeta }{\text{Tr}}\left( {{{\mathbf{W}}_k}\! +\! {{\mathbf{V}}_k}} \right)\! + \!{P_c}} \right) \\
& \!\geqslant\! z,\forall k \in \mathcal{K},\forall j \in \mathcal{J}.
\end{split}
\end{equation}
Then, ${\mathcal{P}_2}$ can be rewritten as
\begin{subequations}\label{eq:p15}
\begin{align}
  {\mathcal{P}_3} \ : &  \mathop {\max }\limits_{{\mathbf{W}},{\mathbf{V}} ,\boldsymbol{\alpha} ,\boldsymbol{\beta},{\boldsymbol{\rho }},{z} } \ \ z \\
   &\ \ {\text{s}}{\text{.t}}{\text{.}} \ \   C1 ,   \ C3, \ C4,  \ C5, \ C6, \ \tilde C7 .
\end{align}
\end{subequations}
On this basis, we divide two steps to solve ${\mathcal{P}_3}$.
%Therefore, the update problem of BF matrix ${\mathbf{W}}$ and AN matrix ${\mathbf{V}}$ can be divided into two steps to solve respectively as follows:

\emph{1) Solve  ${\boldsymbol{\rho }}$ and fix $({{\mathbf{W}},{\mathbf{V}}})$:}  We first fix ${\mathbf{W}}$, ${\mathbf{V}}$, and solve ${\boldsymbol{\rho }}$. Let
\begin{equation}\label{eq:p16}
 {{\partial {f\left( {{\alpha _k^{\left[ {t} \right]}},\beta {_k^j}^{\left[ {t} \right]},\beta {{_k^j}^{\left[ t-1 \right]}},{\rho }_k^j} \right)}} \mathord{\left/
 {\vphantom {{\partial {f\left( {{\alpha _k^{\left[ {t} \right]}},\beta {_k^j}^{\left[ {t} \right]},\beta {{_k^j}^{\left[ t-1 \right]}},{\rho }_k^j} \right)}} {\partial {{\rho }_k^j}}}} \right. \kern-\nulldelimiterspace} {\partial {{\rho }_k^j}}}=0,
 \end{equation}
the optimal ${\rho {_k^j}}^{opt}$ can be obtained as
\begin{equation}\label{eq:p17}
{\rho {_k^j}}^{opt} = \frac{{\sqrt {f\left( {\alpha _k^{\left[ t \right]},\beta{{_k^j}^{\left[ {t} \right]}},\beta {{_k^j}^{\left[ {t - 1} \right]}}} \right)} }}{{\frac{1}{\zeta }{\rm{Tr}}\left( {{\bf{W}}_k^{\left[ t \right]} + {\bf{V}}_k^{\left[ t \right]}} \right) + {P_c}}},\forall k \in \mathcal{K},\forall j \in \mathcal{J}.
\end{equation}

%${\boldsymbol{\alpha}}^{\left[ t \right]}$, ${\boldsymbol{\beta}}^{\left[ t \right]}$, ${\boldsymbol{\delta}}^{\left[ t \right]}$ and ${\boldsymbol{\varsigma}}^{\left[ t \right]}$
%${\boldsymbol{\alpha}}^{\left[ t \right]}$, ${\boldsymbol{\beta}}^{\left[ t \right]}$, ${\boldsymbol{\delta}}^{\left[ t \right]}$ and ${\boldsymbol{\varsigma}}^{\left[ t \right]}$

\emph{2) Solve $({\mathbf{W}},{\mathbf{V}})$ and fix ${\boldsymbol{\rho }}$:} After obtaining ${\boldsymbol{\rho }}$, ${\mathcal{P}_3}$ is still intractable due to the non-convex constraints of $C3$ and $C4$. Here, we introduce auxiliary variable $\boldsymbol{\delta} \in {\mathbb{R}^K}$ with $\boldsymbol{\delta }= \left[ {{\delta _1}, \ldots ,{\delta _K}} \right]$, and $C3$ can be transformed as
\begin{subequations}\label{eq:p18}
\begin{align}
&C8:{\rm{ }}{\alpha _k}{\delta _k} \le {{{\bf{\hat H}}}_k^H}  {{\bf{W}}_k}  {{{\bf{\hat H}}}_k},k \in {\cal K},\\
&C9:{\rm{ }}{\delta _k} \ge {{{\bf{\hat H}}}_k^H}  {{\bf{L}}_k} {{{\bf{\hat H}}}_k} + \sigma _k^2,k \in {\cal K},
\end{align}
\end{subequations}
where ${{{\bf{\hat H}}}_k}= {\bf{H}}_k^H{\hat{\boldsymbol \theta }}^{\left[ t \right]}$. For ${\alpha _k}{\delta _k}$, we can write its upper bound as [33]
\begin{equation}\label{eq:p19}
{\alpha _k}{\delta _k} \le \frac{{\alpha _k^{\left[ t \right]}}}{{2\delta _k^{\left[ t \right]}}}\delta _k^2 + \frac{{\delta _k^{\left[ t \right]}}}{{2\alpha _k^{\left[ t \right]}}}\alpha _k^2,k \in {\cal K}.
\end{equation}
As a result,  $C8$ can be transformed into the following convex constraint
\begin{equation}\label{eq:p20}
\bar C8:\frac{{\alpha _k^{\left[ t \right]}}}{{2\delta _k^{\left[ t \right]}}}\delta _k^2 + \frac{{\delta _k^{\left[ t \right]}}}{{2\alpha _k^{\left[ t \right]}}}\alpha _k^2 \le {{{\bf{\hat H}}}_k^H}  {{\bf{W}}_k} {{{\bf{\hat H}}}_k} ,k \in {\cal K}.
\end{equation}
To deal with $C4$, we introduce two auxiliary variables $\boldsymbol{\varsigma}  \in {\mathbb{R}^{K \times J}}$ with $\boldsymbol{\varsigma}  = \left[ {{\boldsymbol{\varsigma} _1}, \ldots ,{\boldsymbol{\varsigma} _K}} \right]$, ${\boldsymbol{\varsigma} _k} = \left[ {\varsigma _k^1, \ldots ,\varsigma _k^J} \right]$ and $\boldsymbol{\chi}  \in {\mathbb{R}^{K \times J}}$ with $\boldsymbol{\chi}  = \left[ {{\boldsymbol{\chi} _1}, \ldots ,{\boldsymbol{\chi} _K}} \right]$, ${\boldsymbol{\chi} _k} = \left[ {\chi _k^1, \ldots ,\chi _k^J} \right]$, which can be reformulated as
\begin{subequations}\label{eq:p21}
\begin{align}
  &C10:\beta _k^j {{{\bf{\hat H}}}_{e,j}^H} {{\mathbf{L}}_k}  {{{\bf{\hat H}}}_{e,j}} \geqslant {\left( {\varsigma _k^j} \right)^2},\forall k \in \mathcal{K},\forall j \in \mathcal{J},  \\
  &C11:{\left( {\varsigma _k^j} \right)^2} \geqslant \chi_k^j,\forall k \in \mathcal{K},\forall j \in \mathcal{J},\\
  &C12: \chi_k^j \geqslant {{{\bf{\hat H}}}_{e,j}^H}  {{\mathbf{W}}_k} {{{\bf{\hat H}}}_{e,j}}  - \beta _k^j\sigma _{e,j}^2,\forall k \in \mathcal{K},\forall j \in \mathcal{J},
\end{align}
\end{subequations}
where ${{{\bf{\hat H}}}_{e,j}}= {\bf{H}}_{e,j}^H{\hat{\boldsymbol \theta }}^{\left[ t \right]}$. Meanwhile, the non-convex constraint $C10$ can be written as a convex linear matrix inequality as
\begin{equation}\label{eq:p22}
\bar C10: \left[ {\begin{array}{*{23}{c}}
{\beta _k^j}&{{\varsigma}_k^j}\\
{{\varsigma}_k^j}&{{{\bf{\hat H}}}_{e,j}^H}{{\mathbf{L}}_k} {{{\bf{\hat H}}}_{e,j}}
\end{array}} \right]\succeq \textbf{0},\forall k \in \mathcal{K},\forall j \in \mathcal{J}.
\end{equation}
As for  $C11$, we apply the first-order Taylor expansion to ${\left( {\varsigma _k^j} \right)^2}$ for a given point ${\varsigma_k^j}^{\left[ t \right]}$, and it can be approximately translated into ${\left( {\varsigma _k^j} \right)^2} \approx 2{\varsigma_k^j}^{\left[ t \right]}\varsigma _k^j - {\left( {\varsigma {{_k^j}^{\left[ t \right]}}} \right)^2}$. Next, $C11$ can be converted~as
\begin{equation}\label{eq:p23}
\bar C11:2{\varsigma_k^j}^{\left[ t \right]} \varsigma _k^j - {\left( {\varsigma {{_k^j}^{\left[ t \right]}}} \right)^2} \ge \chi_k^j,\forall k \in {\cal K},\forall j \in {\cal J}.
\end{equation}
%Then $C4$ can be replaced by convex constraints $\bar C10$, $\bar C11$ and $C12$. Here, problem ${\mathcal{P}_4}$ can be reformulated when ${\boldsymbol{\rho }}$ is fixed as
Finally, ${\mathcal{P}_3}$ can be formulated as following one
\begin{subequations}\label{eq:p24}
\begin{align}
  {\mathcal{P}_4} \ : &  \mathop {\max }\limits_{{\mathbf{W}},{\mathbf{V}} ,\boldsymbol{\alpha} ,\boldsymbol{\beta},\boldsymbol{\delta},\boldsymbol{\varsigma},\boldsymbol{\chi},{z} } \ \ z \\
   & {\text{s}}{\text{.t}}{\text{.}}  \  C1, C5, C6, \tilde C7, \bar C8, C9, \bar C10, \bar C11, C12.
\end{align}
\end{subequations}
%C1\!,\! \ C5\!,\! \ C6\!,\! \ \tilde C7\!,\! \ \bar C8\!,\! \ C9\!,\!\  \bar C10\!,\! \ \bar C11\!,\! \ C12.

${\mathcal{P}_4}$ is still intractable due to the existence of the rank-one constraints $C5$ and $C6$. Therefore, we choose to drop it and solve its relaxed optimization problem by SDP technique. However, after solving the problem, we need to consider how to convert the obtained ${{\bf{W}}^{opt}}$ and ${{\bf{V}}^{opt}}$ into feasible ${{\bf{w}}^ * }$ and ${{\bf{v}}^ * }$. Specifically, if the optimal solutions ${{\bf{W}}^{opt}}$ and ${{\bf{V}}^{opt}}$ of  ${\mathcal{P}_4}$ satisfy the rank-one condition, we can directly use eigenvalue decomposition to obtain feasible ${{\bf{w}}^ * }$ and ${{\bf{v}}^ * }$. Otherwise, we can obtain an approximate solution by Gaussian randomization technique [34].
%\in {\mathbb{C}^{RN \times RN}}   \in {\mathbb{C}^{\left( {RN + 1} \right) \times \left( {RN + 1} \right)}}

\subsection{Fix $({\mathbf{W}},{\mathbf{V}})$ and Solve $\boldsymbol{\theta}$}
We define ${\bf{Q}} = {\boldsymbol{\theta }}{{\boldsymbol{\theta}}^H} $ and ${\bf{\hat Q}} = {\hat {\boldsymbol{\theta} }}{{{\hat {\boldsymbol{\theta} }}}^H}$,
and then introduce two auxiliary variables ${\boldsymbol{\hat \alpha}} \in {\mathbb{R}^{K \times 1}}$ $\left( \boldsymbol{\hat  \alpha}  = [ {{\hat \alpha _1}, \ldots ,{\hat \alpha _K}} ] \right)$ and $\boldsymbol{\hat  \beta}  \in {\mathbb{R}^{K \times J}}$   $\left( \boldsymbol{\hat  \beta}  = [ {{\boldsymbol{\hat  \beta} _1}, \ldots ,{\boldsymbol{\hat  \beta} _K}} ] \right.$, $\left.{\boldsymbol{\hat  \beta} _k} = [ {{\hat  \beta _{k}^1}, \ldots ,{\hat  \beta _{k}^J}} ]\right)$. Similarly, for given ${{\bf{W}}^{\left[ t \right]}}$ and ${{\bf{V}}^{\left[ t \right]}}$,  ${\mathcal{P}_0}$ can be reformulated as
\begin{subequations}\label{eq:p25}
\begin{align}
  {\mathcal{P}_5} \  &: \mathop {\max }\limits_{ {\bf{\hat Q}} ,\boldsymbol{\hat \alpha} ,\boldsymbol{\hat \beta},{z} } \ z \\
   {\text{s}}{\text{.t}}{\text{.}}  \ &{\bar C3}:  {{\hat \alpha} _k} \le \frac{{{\rm{Tr}}\left( {{\mathbf{\bar H}}_k^W {\bf{\hat Q}}} \right)}}{{{\rm{Tr}}\left( {{\mathbf{\bar H}}_k^L{\bf{\hat Q}}} \right) + \sigma _k^2}},  \forall k \in {\cal K},  \\
  &{\bar C4}:   {\hat \beta}_k^j \ge \frac{{{\rm{Tr}}\left( {{\mathbf{\bar H}}_{k,j}^W {\bf{\hat Q}}} \right)}}{{{\rm{Tr}}\left( {{\mathbf{\bar H}}_{k,j}^L {\bf{\hat Q}}} \right) + \sigma _{e,j}^2}},\forall k \in {\cal K},\forall j \in {\cal J},\\
   &{ C13}: {\rm{Tr}}\left( {{{\bf{E}}_m}{\bf{\hat Q}}} \right) = 1, {\bf{\hat Q}} \succeq 0,m = 1,...,RN + 1,\\
  & { C14}:  {\rm{rank}}\left( {{\bf{\hat Q}}} \right) = 1,\\
   & \bar C7  ,
\end{align}
\end{subequations}
where ${{\mathbf{E}}_m} \in {\mathbb{R}^{\left( {RN + 1} \right) \times \left( {RN + 1} \right)}}$ satisfies that the element at position $\left( {m,m} \right)$ is 1, otherwise it is 0, ${\mathbf{\bar H}}_k^W = {{\mathbf{H}}_k}{{\mathbf{W}}_k^{\left[ t \right]}}{\mathbf{H}}_k^H$, ${\mathbf{\bar H}}_k^L = {{\mathbf{H}}_k}{{\mathbf{L}}_k^{\left[ t \right]}}{\mathbf{H}}_k^H$, ${\mathbf{\bar H}}_{k,j}^W = {{\mathbf{H}}_{e,j}}{{\mathbf{W}}_k^{\left[ t \right]}}{\mathbf{H}}_{e,j}^H$, and ${\mathbf{\bar H}}_{k,j}^L = {{\mathbf{H}}_{e,j}}{{\mathbf{L}}_k^{\left[ t \right]}}{\mathbf{H}}_{e,j}^H$.  We can clearly observe that ${\mathcal{P}_5}$ is intractable due to the existence of constraints ${\bar C3}$ and ${\bar C4}$. Similar  to deal with  $C3$ and $C4$ in Section III-B, we can obtain
\begin{subequations}\label{eq:p26}
\begin{align}
  {\mathcal{P}_6} \  &: \mathop {\max }\limits_{ {\bf{\hat Q}} ,\boldsymbol{\hat \alpha} ,\boldsymbol{\hat \beta}, \boldsymbol{\delta},\boldsymbol{\varsigma},\boldsymbol{\chi},{z} } \ z \\
   {\text{s}}{\text{.t}}{\text{.}} \  & \tilde C8:\frac{{{\hat \alpha}_k^{\left[ t \right]}}}{{2\delta _k^{\left[ t \right]}}}\delta _k^2 + \frac{{\delta _k^{\left[ t \right]}}}{{2{\hat \alpha}_k^{\left[ t \right]}}}{\hat \alpha}_k^2 \leqslant {\text{Tr}}\left( {{\mathbf{\bar H}}_k^W{\mathbf{\hat Q}}} \right),k \in \mathcal{K},   \\
 & \tilde C9:{\delta _k} \geqslant {\text{Tr}}\left( {{\mathbf{\bar H}}_k^L{\mathbf{Q}}} \right) + \sigma _k^2,k \in \mathcal{K}, \\
 & \tilde C10: \left[ {\begin{array}{*{23}{c}}
  {{\hat \beta} _k^j}&{{\varsigma}_k^j}\\
  {{\varsigma}_k^j}&{{\text{Tr}}\left( {{\mathbf{\bar H}}_{k,j}^L{\mathbf{\hat Q}}} \right)}
  \end{array}} \right]\!\succeq\! \textbf{0},\forall k \in \mathcal{K}\!,\!\forall j \in \mathcal{J}\!,\! \\
 & \tilde C12: \chi_k^j\! \geqslant\! {\text{Tr}}\left( {{\mathbf{\bar H}}_{k,j}^W{\mathbf{\hat Q}}} \right)\!-\! {\hat \beta}_k^j\sigma _{e,j}^2,\forall k \in \mathcal{K}\!,\!\forall j \in \mathcal{J}\!,\! \\
  &  \bar C7 ,\ \bar C11,  \ C13,\ C14 .
\end{align}
\end{subequations}

\setcounter{TempEqCnt}{\value{equation}}
\setcounter{equation}{28}
\begin{figure*}[hb]
\hrulefill
%\vspace*{10pt}
\begin{equation}\label{eq:p29}
\begin{split}
& \Pr \left\{ {{{\log }_2}\left( {1 + \frac{{{{\left| {{{\hat {\boldsymbol{\theta} }}^H}{{\bf{H}}_{e,j}}{{\bf{w}}_k}} \right|}^2}}}{{\sum\limits_{i = 1,i \ne k}^K {{{\left| {{{\hat {\boldsymbol{\theta} }}^H}{{\bf{H}}_{e,j}}{{\bf{w}}_i}} \right|}^2} + \sum\limits_{i = 1}^K {{{\left| {{{\hat {\boldsymbol{\theta} }}^H}{{\bf{H}}_{e,j}}{{\bf{v}}_i}} \right|}^2} + } \sigma _{e,j}^2} }}} \right) \le R_k^{re}} \right\}\\
 =&  \Pr \left\{ {\left( {{2^{R_k^{re}}} - 1} \right)\left( {\sum\limits_{i = 1,i \ne k}^K {{{\left| {{{\hat {\boldsymbol{\theta} }}^H}{{\bf{H}}_{e,j}}{{\bf{w}}_i}} \right|}^2} + \sum\limits_{i = 1}^K {{{\left| {{{\hat {\boldsymbol{\theta} }}^H}{{\bf{H}}_{e,j}}{{\bf{v}}_i}} \right|}^2} + } \sigma _{e,j}^2} } \right) - {{\left| {{{\hat {\boldsymbol{\theta} }}^H}{{\bf{H}}_{e,j}}{{\bf{w}}_k}} \right|}^2} \ge 0} \right\}\\
 =&  \Pr \left\{ {\left( {{\bf{h}}_{d,e,j}^H + {\bf{f}}_{e,j}^H{{\bf{\Theta }}^H}{\bf{G}}} \right){{\bf{D}}_k}{{\left( {{\bf{h}}_{d,e,j}^H + {\bf{f}}_{e,j}^H{{\bf{\Theta }}^H}{\bf{G}}} \right)}^H} + \left( {{2^{R_k^{re}}} - 1} \right)\sigma _{e,j}^2 \ge 0} \right\}  \\
  =&  \Pr \left\{ {{\bf{\tilde h}}_j^H{{\bf{D}}_k}{{{\bf{\tilde h}}}_j} + {\bf{\tilde h}}_j^H{{\bf{D}}_k}\Delta {{\bf{h}}_j} + \Delta {\bf{h}}_j^H{{\bf{D}}_k}{{{\bf{\tilde h}}}_j} + \Delta {\bf{h}}_j^H{{\bf{D}}_k}\Delta {{\bf{h}}_j} + \tilde \sigma _j^2 \ge 0} \right\}.
\end{split}
\end{equation}
\end{figure*}
\setcounter{equation}{\value{TempEqCnt}}

\begin{algorithm}[t]
	\renewcommand{\algorithmicrequire}{\textbf{Input:}}
	\renewcommand{\algorithmicensure}{\textbf{Output:}}
	\caption{Proposed Algorithm Based on Perfect CSI.}
	\label{Algorithm1}
	\begin{algorithmic}[1]
		\REQUIRE ${{\bf{h}}_{b,k}}$, ${{\bf{h}}_{b,e,j}}$, ${{\bf{f}}_{r,k}}$, ${{\bf{f}}_{r,e,j}}$, ${{\bf{G}}_{b,r}}$, and ${\sigma _k}$, ${\sigma _e}$.
		\ENSURE Beamforming matrix ${\bf{W}}$, AN matrix ${\bf{V}}$, phase shift matrix ${\bf{\hat Q}}$, and SEE $z$.
		\STATE Initialize ${\bf{W}}^{\left[ t \right]}$, ${\bf{V}}^{\left[ t \right]}$, ${\bf{\hat Q}}^{\left[ t \right]}$,  ${\boldsymbol{\alpha}}^{\left[ n \right]}$, ${\boldsymbol{\beta}}^{\left[ t \right]}$, ${\boldsymbol{ \beta}}^{\left[ t-1 \right]}$, ${\boldsymbol{\delta}}^{\left[ t \right]}$, ${\boldsymbol{\varsigma}}^{\left[ t \right]}$, ${z^{\left[ t \right]}}$, $t=0$, and threshold $\tau$;
		\WHILE{ ${z^{\left[ t \right]}}-{z^{\left[ t-1 \right]}}>{\tau}$}
		\STATE $t=t+1$;
		\STATE Update ${\boldsymbol{\rho }}^{\left[ t \right]}$ by (17);
		\STATE Update ${\bf{W}}^{\left[ t \right]}$, ${\bf{V}}^{\left[ t \right]}$, by solving ${{\cal P}_4}$;
		\STATE Update $z^{\left[ t \right]}$, ${\bf{\hat Q}}^{\left[ t \right]}$, by solving ${{\cal P}_6}$;
		\ENDWHILE
		\STATE \textbf{return} ${\bf{W}}^{\left[ t \right]}$, ${\bf{V}}^{\left[ t \right]}$, ${\bf{\hat Q}}^{\left[ t \right]}$, and $z^{\left[ t \right]}$.
	\end{algorithmic}
\end{algorithm}

${\mathcal{P}_6}$ is still difficult to solve due to the rank-one constraint. Similar to the solutions of $\mathbf{W}$ and $\mathbf{V}$ for the rank-one constraints in Section III-A, we use the SDP technique to solve the above optimization problem. Meanwhile, we can use the Gaussian randomization method to obtain the approximate solution of $\boldsymbol{\theta}^{*}$ when it does not satisfy the rank-one constraint.

Summarily, to solve ${\mathcal{P}_0}$, we first initialize the feasible points ${{\bf{W}}^{\left[ t \right]}}$, ${{\bf{V}}^{\left[ t \right]}}$, ${\bf{\hat Q}}^{\left[ t \right]}$, ${\boldsymbol{\alpha}}^{\left[ t \right]}$, ${\boldsymbol{\beta}}^{\left[ t \right]}$, ${\boldsymbol{\beta}}^{\left[ t-1 \right]}$, ${\boldsymbol{\delta}}^{\left[ t \right]}$ and ${\boldsymbol{\varsigma}}^{\left[ t \right]}$, and then solve ${\boldsymbol{\rho }}$ by (17), and solve  ${\mathcal{P}_4}$ to obtain ${{\bf{W}}^{\left[ t+1 \right]}}$, ${{\bf{V}}^{\left[ t+1 \right]}}$, and solve ${\mathcal{P}_6}$ to obtain $z^{\left[ t+1 \right]}$, ${\bf{\hat Q}}^{\left[ t+1 \right]}$, and then repeat (17), ${\mathcal{P}_4}$ and ${\mathcal{P}_6}$ until the result $z$ converges to a stable value, which is summarized as \textbf{Algorithm 1}.
%Due to the removal of the rank-one constraints of ${\bf{W}}$ and ${\bf{\hat Q}}$, the entire optimization process is a convex optimization process, and the iterative update will maintain the non-reducing value of $z$ each time until it converges to a locally stable solution.

\section{Extend to Imperfect CSI for the Eves' links}
In practical communication environment, the BS may reliably estimate the CSI of legitimate users' links with negligible estimation error. However, the passive nature of Eves  makes the BS difficult to accurately estimate the CSI for the Eves' links [35], [36]. In this section, we consider imperfect CSI for the Eves' links, which are expressed as follows:
\begin{equation}\label{eq:27}
\begin{split}
&{{\bf{h}}_{d,e,j}} = {{{\bf{\tilde h}}}_{d,e,j}} + \Delta {{\bf{h}}_{d,e,j}},\forall k \in \mathcal{K},\forall j \in \mathcal{J},\\
&{{\bf{f}}_{e,j}} = {{{\bf{\tilde f}}}_{e,j}} + \Delta {{\bf{f}}_{e,j}},\forall k \in \mathcal{K},\forall j \in \mathcal{J},
\end{split}
\end{equation}
where ${{{\bf{\tilde h}}}_{d,e,j}}$ and ${{{\bf{\tilde f}}}_{e,j}}$ represent the corresponding actual estimation values, $\Delta {{\bf{h}}_{d,e,j}}$ and $\Delta {{\bf{f}}_{e,j}}$ represent the corresponding estimation errors and satisfy $\Delta {{\bf{h}}_{d,e,j}} \sim {{\cal D}^D} =  {\cal C}{\cal N}\left( {{\bf{0}},{{\bf{E}}_{d,j}}} \right)$ with ${{\bf{E}}_{d,j}} = \sigma _{d,j}^2{{\bf{I}}_{MB}}$, and ${\rm{ }}\Delta {{\bf{f}}_{e,j}} \sim  {{\cal D}^R} =  {\cal C}{\cal N}\left( {{\bf{0}},{{\bf{E}}_{f,j}}} \right)$ with ${{\bf{E}}_{f,j}} = \sigma _{f,j}^2{{\bf{I}}_{RN}}$.
We define ${R_{k,j}}$ as the achievable rate of the $j$-th Eve eavesdropping on the $k$-th user. When ${R_{k,j}}$ exceeds the redundancy $R_k^{re}$, a  secrecy outage event of the $k$-th user will occur at the BS [37]. Based on this, ${\mathcal{P}_0}$ can be converted to the following one
\begin{subequations}\label{eq:p28}
\begin{align}
  {{\hat {\cal P}}_0}   & : \mathop {\max }\limits_{{\mathbf{W}},{\mathbf{V}},\boldsymbol{\theta}} \;  \mathop {\min }\limits_{k \in {\cal K}} \ \  \eta _k^{EE} \\
  {\text{s}}{\text{.t}}{\text{.}}  \;& C15:  {\rm{Pr}}\left\{ {{{\log }_2}\left( {1 + {\gamma _{k,j}}} \right) \le R_k^{re}} \right\} \ge 1 - {\varphi _k},\forall k \in \mathcal{K}, \nonumber \\
  & \qquad \qquad \qquad \qquad \qquad \qquad \qquad \qquad   \quad  \forall j \in \mathcal{J},\\
  &  C16:  \Delta {{\bf{h}}_{d,e,j}} \in {{\cal D}^D},\Delta {{\bf{f}}_{e,j}} \in {{\cal D}^R}, \forall j \in {\cal J},\\
   &   C1  ,  \ C2,
\end{align}
\end{subequations}
where ${\varphi _k} \in \left( {0,1} \right)$ denotes the maximum outage probability. %, and $C16$ denotes the set of uncertain channel for the Eves' links.
As stated before, solving the  optimization problem is extremely challenging due to the non-convexity of the objective function and the new constraints $C15$ and $C16$.

We first deal with $C15$, which can be simplified to (29), where ${{\bf{D}}_k} = \sum\limits_{i = 1,i \ne k}^K {\left( {{2^{R_k^{re}}} - 1} \right){{\bf{W}}_i} + \sum\limits_{i = 1}^K {\left( {{2^{R_k^{re}}} - 1} \right){{\bf{V}}_i}} }  - {{\bf{W}}_k}$, ${{{\bf{\tilde h}}}_j} = {{{\bf{\tilde h}}}_{d,e,j}} + {{\bf{G}}^H}{\bf{\Theta }}{{{\bf{\tilde f}}}_{e,j}}$, $\Delta {{\bf{h}}_j} = \Delta {{\bf{h}}_{d,e,j}} + {{\bf{G}}^H}{\bf{\Theta }}\Delta {{\bf{f}}_{e,j}}$, and $\tilde \sigma _j^2 = \left( {{2^{R_k^{re}}} - 1} \right)\sigma _{e,j}^2$. However, the above transformation is still difficult to address. Fortunately, the interruption constraint can be handled with the BTI of Lemma 1 [38], i.e.

\setcounter{TempEqCnt}{\value{equation}}
\setcounter{equation}{46}
\begin{figure*}[hb]
\hrulefill
\begin{equation}\label{eq:p47}
\begin{split}
C19: \left[ {\begin{array}{*{20}{c}}
  {\kappa _k^j{{\mathbf{I}}_{(MB + RN)}} - {{\mathbf{C}}_{W,k}}}&{ - {\mathbf{C}}_{W,k}^H{{{\mathbf{\tilde x}}}_j}} \\
  { - {\mathbf{\tilde x}}_j^H{{\mathbf{C}}_{W,k}}}&{ {- \kappa _k^j} {\psi _j^2 {\nu}  \left( {{\mathbf{Tr}}\left( {{{\mathbf{E}}_{d,j}}} \right) + {\mathbf{Tr}}\left( {{{\mathbf{E}}_{f,j}}} \right)} \right)}  - {\mathbf{\tilde x}}_j^H{{{\mathbf{C}}_{W,k}}}{{{\mathbf{\tilde x}}}_j} + \varpi _k^j}
\end{array}} \right] \succeq {\mathbf{0}},\forall k \in \mathcal{K}, j \in \mathcal{J}, \qquad \
\end{split}
\end{equation}
\end{figure*}
\setcounter{equation}{\value{TempEqCnt}}

\setcounter{TempEqCnt}{\value{equation}}
\setcounter{equation}{49}
\begin{figure*}[hb]
%\hrulefill
\begin{equation}\label{eq:p50}
\begin{split}
C20: \left[ {\begin{array}{*{20}{c}}
  {\omega _k^j{{\mathbf{I}}_{\left( {MB + RN} \right)}} + {{\mathbf{C}}_{L,k}}}&{{\mathbf{C}}_{L,k}^H{{{\mathbf{\tilde x}}}_j}} \\
  {{\mathbf{\tilde x}}_j^H{{\mathbf{C}}_{L,k}}}&{ - \omega _k^j\psi _j^2\nu \left( {{\mathbf{Tr}}\left( {{{\mathbf{E}}_{d,j}}} \right) + {\mathbf{Tr}}\left( {{{\mathbf{E}}_{f,j}}} \right)} \right) + {\mathbf{\tilde x}}_j^H{{\mathbf{C}}_{L,k}}{{{\mathbf{\tilde x}}}_j} - \chi _k^j + \sigma _{e,j}^2}
\end{array}} \right] \succeq{\mathbf{0}},\forall k \in \mathcal{K}, j \in \mathcal{J},
\end{split}
\end{equation}
\end{figure*}
\setcounter{equation}{\value{TempEqCnt}}

\emph{Lemma 1 (BTI):} Assume a probability constraint
\setcounter{equation}{29}
\begin{equation}\label{eq:p30}
f\left( {\bf{x}} \right) = \Pr \left\{ {{{\bf{x}}^H}{\bf{Ax}} + 2{\mathop{\rm Re}\nolimits} \left\{ {{{\bf{u}}^H}{\bf{x}}} \right\} + c \ge 0} \right\} \ge 1 - \varphi ,
\end{equation}
where ${\mathbf{A}} \in {\mathbb{H}^{P \times P}}$, ${\mathbf{u}} \in {\mathbb{C}^{P \times 1}}$,  ${\mathbf{x}} \in {\mathbb{C}^{P \times 1}} \sim \mathcal{C}\mathcal{N}\left( {{\mathbf{0}},{\mathbf{I}}} \right)$, and $\varphi   \in \left( {\left. {0,1} \right]} \right.$. By introducing two slack variables $\lambda$ and $\varepsilon$, the following relationship always holds
\begin{equation}\label{eq:p31}
\begin{split}
\left\{ \begin{gathered}
  \ {\text{Tr}}\left( {\mathbf{A}} \right) - \sqrt { - 2\ln \left( \varphi  \right)} \lambda  + \ln \left( \varphi  \right)\varepsilon  + c \geqslant 0, \hfill \\
 \  \left\| {\left[ {\begin{array}{*{20}{c}}
  {{\text{vec}}\left( {\mathbf{A}} \right)} \\
  {\sqrt 2 {\mathbf{u}}}
\end{array}} \right]} \right\| \leqslant \lambda , \hfill \\
 \  \varepsilon {\mathbf{I}} + {\mathbf{A}} \succeq \mathbf{ 0},\varepsilon  \geqslant 0. \hfill \\
\end{gathered}  \right.
\end{split}
\end{equation}
We define $\Delta {{{\mathbf{\tilde h}}}_{d,e,j}} = {\sigma _{d,j}}\Delta {{\mathbf{e}}_{d,j}}$ and $\Delta {{{\mathbf{\tilde f}}}_{e,j}} = {\sigma _{f,j}}\Delta {{\mathbf{e}}_{f,j}}$, where $\Delta {{\mathbf{e}}_{d,j}} \sim \mathcal{C}\mathcal{N}\left( {{\mathbf{0}},{\mathbf{I}}} \right)$ and $\Delta {{\mathbf{e}}_{f,j}} \sim \mathcal{C}\mathcal{N}\left( {{\mathbf{0}},{\mathbf{I}}} \right)$. By citing Lemma 1, (30) becomes
\begin{equation}\label{eq:p32}
\begin{split}
\Pr \left\{ {\Delta {\mathbf{e}}_j^H{{\mathbf{A}}_k}\Delta {{\mathbf{e}}_j} + 2\operatorname{Re} \left\{ {{\mathbf{u}}_{k,j}^H\Delta {{\mathbf{e}}_j}} \right\} + {c_{k,j}} \geqslant 0} \right\} \geqslant 1 - {\varphi _k},
\end{split}
\end{equation}
where $ \Delta {\mathbf{x}} = \left[ {\Delta {{\mathbf{h}}_{d,e,j}};\Delta {{\mathbf{f}}_{e,j}}} \right]$, ${c_{k,j}} = {\mathbf{\tilde h}}_j^H{{\mathbf{D}}_k}{{{\mathbf{\tilde h}}}_j} + \tilde \sigma _j^2$,
\begin{equation}\label{eq:p33}
\begin{split}
 &{{\mathbf{A}}_k} = \left[ {\begin{array}{*{20}{c}}
  {\sigma _{d,j}^2{{\mathbf{D}}_k}}&{{\sigma _{d,j}}{\sigma _{f,j}}{{\mathbf{D}}_k}{{\mathbf{G}}^H}{\mathbf{\Theta }}} \\
  {{\sigma _{d,j}}{\sigma _{f,j}}{{\mathbf{\Theta }}^H}{\mathbf{G}}{{\mathbf{D}}_k}}&{\sigma _{f,j}^2{{\mathbf{\Theta }}^H}{\mathbf{G}}{{\mathbf{D}}_k}{{\mathbf{G}}^H}{\mathbf{\Theta }}}
\end{array}} \right], \\
& {{\mathbf{u}}_{k,j}} = {\left[ {{{\left( {{\sigma _{d,j}}{{\mathbf{D}}_k}{{{\mathbf{\tilde h}}}_j}} \right)}^H},{{\left( {{\sigma _{f,j}}{{\mathbf{\Theta }}^H}{\mathbf{G}}{{\mathbf{D}}_k}{{{\mathbf{\tilde h}}}_j}} \right)}^H}} \right]^H}.
\end{split}
\end{equation}
We introduce two auxiliary variables  $\boldsymbol{\lambda}  \in {\mathbb{R}^{K \times J}}$ with $\boldsymbol{\lambda}  = \left[ {{\boldsymbol{\lambda} _1}, \ldots ,{\boldsymbol{\lambda} _K}} \right]$, ${\boldsymbol{\lambda} _k} = \left[ {\lambda _k^1, \ldots ,\lambda _k^J} \right]$ and $\boldsymbol{\varepsilon}  \in {\mathbb{R}^{K \times J}}$ with $\boldsymbol{\varepsilon}  = \left[ {{\boldsymbol{\varepsilon} _1}, \ldots ,{\boldsymbol{\varepsilon} _K}} \right]$, ${\boldsymbol{\varepsilon} _k} = \left[ {\varepsilon_k^1, \ldots ,\varepsilon _k^J} \right]$, then (31) can be converted as
\begin{equation}\label{eq:p34}
\begin{split}
\bar C15:\left\{ \begin{gathered}
  {\text{Tr}}\left( {{{\mathbf{A}}_k}} \right) - \sqrt { - 2\ln \left( {{\varphi _k}} \right)} \lambda _k^j + \ln \left( {{\varphi _k}} \right)\varepsilon _k^j + {c_{k,j}} \geqslant 0, \hfill \\
  \left\| {\left[ {\begin{array}{*{20}{c}}
  {{\text{vec}}\left( {{{\mathbf{A}}_k}} \right)} \\
  {\sqrt 2 {{\mathbf{u}}_{k,j}}}
\end{array}} \right]} \right\| \leqslant \lambda _k^j, \hfill \\
  \varepsilon _k^j{\mathbf{I}} + {{\mathbf{A}}_k} \succeq {\mathbf{0}},\varepsilon _k^j \geqslant 0, \hfill \\
\end{gathered}  \right.
\end{split}
\end{equation}
where $\forall k \in {\cal K}$ and $\forall j \in {\cal J}$. $C2$, $C3$, $C4$, $C7$, and $\bar C15$ are still non-convex and intractable due to the coupling of ${\mathbf{W}}$, ${\mathbf{V}}$ and $\boldsymbol{\theta}$ as well as the constraint of the unit modulo. Next, we propose an alternative scheme to deal with it.

\setcounter{TempEqCnt}{\value{equation}}
\setcounter{equation}{53}
\begin{figure*}[hb]
\hrulefill
\begin{equation}\label{eq:p54}
\begin{split}
 & {{{\mathbf{\hat A}}}_k} = \left[ {\begin{array}{*{20}{c}}
  {\sigma _{d,j}^2{{\mathbf{D}}_k}} \ & \ {{\sigma _{d,j}}{\sigma _{f,j}}{{\mathbf{D}}_k}{{\mathbf{G}}^H}{{\text{diag}}\left( {{{[ {{\mathbf{\hat Q}}} ]}_{1:RN,RN + 1}}} \right)}} \\
  {{\sigma _{d,j}}{\sigma _{f,j}}{\left({{\text{diag}}\left( {{{[ {{\mathbf{\hat Q}}} ]}_{1:RN,RN + 1}}} \right)}\right)^H}{\mathbf{G}}{{\mathbf{D}}_k}} \ & \ {\sigma _{f,j}^2\sum\nolimits_s {{x_{k,s}}} {{\mathbf{O}}_{k,s}}{\mathbf{\hat Q}}{{\mathbf{V}}_{k,s}}}
\end{array}} \right],
\end{split}
\end{equation}
\end{figure*}
\setcounter{equation}{\value{TempEqCnt}}

\setcounter{TempEqCnt}{\value{equation}}
\setcounter{equation}{56}
\begin{figure*}[hb]
%\hrulefill
\begin{subequations}\label{eq:p57}
\begin{align}
& {{\mathbf{\hat C}}_{W,k}} = \left[ {\begin{array}{*{20}{c}}
  {{{\mathbf{W}}_k}} \quad & \quad {{\mathbf{W}}_k}{{\mathbf{G}}^H}{{\text{diag}}\left( {{{[ {{\mathbf{\hat Q}}} ]}_{1:RN,RN + 1}}} \right)} \\
  {\left({{\text{diag}}\left( {{{[ {{\mathbf{\hat Q}}} ]}_{1:RN,RN + 1}}} \right)}\right)^H}{\mathbf{G}}{{\mathbf{W}}_k} \quad & \quad \sum\nolimits_s {{{\hat x}_{k,s}}} {{{\mathbf{\hat O}}}_{k,s}}{\mathbf{\hat Q}}{{{\mathbf{\hat V}}}_{k,s}}
\end{array}} \right].\\
& {{\mathbf{\hat C}}_{L,k}} = \left[ {\begin{array}{*{20}{c}}
  {{{\mathbf{L}}_k}} \quad & \quad {{\mathbf{L}}_k}{{\mathbf{G}}^H}{{\text{diag}}\left( {{{[ {{\mathbf{\hat Q}}} ]}_{1:RN,RN + 1}}} \right)} \\
  {\left({{\text{diag}}\left( {{{[ {{\mathbf{\hat Q}}} ]}_{1:RN,RN + 1}}} \right)}\right)^H}{\mathbf{G}}{{\mathbf{L}}_k} \quad & \quad \sum\nolimits_s {{{\tilde x}_{k,s}}} {{{\mathbf{\tilde O}}}_{k,s}}{\mathbf{\hat Q}}{{{\mathbf{\tilde V}}}_{k,s}}
\end{array}} \right].
\end{align}
\end{subequations}
\end{figure*}
\setcounter{equation}{\value{TempEqCnt}}

\subsection{Fix ${\boldsymbol{\theta }}$ and Solve $({\mathbf{W}},{\mathbf{V}})$}
%We can observe that only the constraints related to the Eves' channel are changed for problems ${{\hat {\cal P}}_0}$ and ${{{\cal P}}_0}$. Therefore,
Similar to the above scheme, when ${\boldsymbol{\theta }}$ is fixed, ${{\hat {\cal P}}_0}$ can be transformed into
\begin{subequations}\label{eq:p35}
\begin{align}
 & {{\hat {\cal P}}_3}  :   \mathop {\max }\limits_{{\mathbf{W}},{\mathbf{V}} ,\boldsymbol{\alpha} ,\boldsymbol{\beta},{\boldsymbol{\rho }},{\boldsymbol{\lambda}},{\boldsymbol{\varepsilon}},{z} } \ \ z \\
   & {\text{s}}{\text{.t}}{\text{.}} \ {\tilde C4}: \beta _k^j \geqslant \frac{{{{\left( {{{{\mathbf{\tilde h}}}_j} + \Delta {{\mathbf{h}}_j}} \right)}^H}{{\mathbf{W}}_k}\left( {{{{\mathbf{\tilde h}}}_j} + \Delta {{\mathbf{h}}_j}} \right)}}{{{{\left( {{{{\mathbf{\tilde h}}}_j} + \Delta {{\mathbf{h}}_j}} \right)}^H}{{\mathbf{L}}_k}\left( {{{{\mathbf{\tilde h}}}_j} + \Delta {{\mathbf{h}}_j}} \right) + \sigma _{e,j}^2}}, \nonumber  \\
   & \qquad \qquad \qquad \qquad \qquad \qquad \qquad \quad  \forall k \in \mathcal{K}, \forall j \in \mathcal{J},\\
   & \quad  \  \;  C1 ,   \;  C3, \;   \; C5, \; C6, \; \tilde C7, \;  \bar C15, \; C16.
\end{align}
\end{subequations}
The optimal ${\boldsymbol{\rho }}^{opt}$ can be obtained by using (18), and $C3$ can be transformed into $\bar C8$ and $C9$. Next,  we introduce three auxiliary variables $\boldsymbol{\varsigma}  \in {\mathbb{R}^{K \times J}}$ with $\boldsymbol{\varsigma}  = \left[ {{\boldsymbol{\varsigma} _1}, \ldots ,{\boldsymbol{\varsigma} _K}} \right]$, ${\boldsymbol{\varsigma} _k} = \left[ {\varsigma _k^1, \ldots ,\varsigma _k^J} \right]$,  $\boldsymbol{\chi}  \in {\mathbb{R}^{K \times J}}$ with $\boldsymbol{\chi}  = \left[ {{\boldsymbol{\chi} _1}, \ldots ,{\boldsymbol{\chi} _K}} \right]$, ${\boldsymbol{\chi} _k} = \left[ {\chi _k^1, \ldots ,\chi _k^J} \right]$, and $\boldsymbol{\varpi}  \in {\mathbb{R}^{K \times J}}$  with $\boldsymbol{\varpi}  = \left[ {{\boldsymbol{\varpi} _1}, \ldots ,{\boldsymbol{\varpi} _K}} \right]$, ${\boldsymbol{\varpi} _k} = \left[ {\varpi _k^1, \ldots ,\varpi _k^J} \right]$,
and thus, ${\tilde C4}$ can be transformed into the following constraints
\begin{subequations}\label{eq:p36}
\begin{align}
  &\ {\left( {{{{\mathbf{\tilde h}}}_j} + \Delta {{\mathbf{h}}_j}} \right)^H}{{\mathbf{W}}_k}\left( {{{{\mathbf{\tilde h}}}_j} + \Delta {{\mathbf{h}}_j}} \right) \leqslant \varpi _k^j,\forall k \in \mathcal{K}, j \in \mathcal{J},  \\
  &\ \varpi _k^j \leqslant {\left( {\varsigma _k^j} \right)^2},\forall k \in \mathcal{K}, \forall j \in \mathcal{J},  \\
 &\ {\left( {\varsigma _k^j} \right)^2} \leqslant \beta _k^j\chi _k^j,\forall k \in \mathcal{K},  \forall j \in \mathcal{J},  \\
 &\  \chi _k^j \leqslant {\left( {{{{\mathbf{\tilde h}}}_j} + \Delta {{\mathbf{h}}_j}} \right)^H}{{\mathbf{L}}_k}\left( {{{{\mathbf{\tilde h}}}_j} + \Delta {{\mathbf{h}}_j}} \right) + \sigma _{e,j}^2, \forall k \in \mathcal{K},  \nonumber  \\
 & \qquad \qquad \qquad \qquad \qquad \qquad \qquad \qquad \qquad  \forall \  j \in \mathcal{J}.
\end{align}
\end{subequations}
Consequently, we can transform (36b) and (36c) into convex constraints as follows:
\begin{subequations}\label{eq:p37}
\begin{align}
  &C17:  \varpi _k^j \leqslant 2{{\varsigma _k^j}^{\left[ t \right]}}\varsigma _k^j - {\left( {\varsigma {{_k^j}^{\left[ t \right]}}} \right)^2}{\text{,}}\forall k \in \mathcal{K}{\text{,}}\forall j \in \mathcal{J}{\text{,}}  \\
 &C18:  \left[ {\begin{array}{*{20}{c}}
  {\beta _k^j}&{\varsigma _k^j} \\
  {\varsigma _k^j}&{\chi _k^j}
\end{array}} \right] \succeq {\mathbf{0}},\forall k \in \mathcal{K}{\text{,}}\forall j \in \mathcal{J}.
\end{align}
\end{subequations}
Next, we observe that (36a) and (36b) are non-probabilistic constraints that cannot be handled with BTI. For convenience, we define the Eves' equivalent estimation channel and estimation error vectors as
\begin{equation}\label{eq:p38}
\begin{split}
\left\{ \begin{gathered}
  {{{\mathbf{\tilde x}}}_j} = {\left[ {{\mathbf{\tilde h}}_{d,e,j}^H,{\mathbf{\tilde f}}_{e,j}^H} \right]^H}, \hfill \\
  \Delta {{\mathbf{x}}_j} = {\left[ {\Delta {\mathbf{h}}_{d,e,j}^H,\Delta {\mathbf{f}}_{e,j}^H} \right]^H}. \hfill \\
\end{gathered}  \right.
\end{split}
\end{equation}
To remove the estimation error, we adopt the Sphere Boundary method [39] and $C15$ can be rewritten as
\begin{equation}\label{eq:p39}
\begin{split}
& {\mathbf{e}}_j^H{{\mathbf{A}}_k}\Delta {{\mathbf{e}}_j} + 2\operatorname{Re} \left\{ {{\mathbf{u}}_{k,j}^H\Delta {{\mathbf{e}}_j}} \right\} + {c_{k,j}} \geqslant 0,  \\
 & \forall \Delta {\mathbf{e}}_j^H\Delta {{\mathbf{e}}_j} \leqslant \psi _k^2, \forall k \in \mathcal{K}{\text{,}}\forall j \in \mathcal{J},
 \end{split}
\end{equation}
where the Gaussian random vector ${{\mathbf{e}}_j}$ satisfying
\begin{equation}\label{eq:p40}
\mathcal{S} = \left\{ {\Delta {{\mathbf{e}}_j}|\Pr \left\{ {\Delta {\mathbf{e}}_j^H\Delta {{\mathbf{e}}_j} \leqslant {\psi _k}} \right\} =  1 - {\varphi _k}} \right\},
\end{equation}
%Here, the uncertainty region radius $\psi _j$ consists of $\psi _j^2 = \psi _{d,j}^2 + \psi _{f,j}^2$ and satisfies F.
and the region radius $\psi _k$ satisfying
\begin{equation}\label{eq:p41}
\psi _k = \sqrt {\frac{1}{2}F_{\chi _{2\left( {MB + RN} \right)}^2}^{ - 1}\left( {1 - {\varphi _k}} \right)},\forall k \in \mathcal{K},\ \ \
\end{equation}
where $F_{\chi _{2\left( {MB + RN} \right)}^2}^{ - 1}\left( {1 - {\varphi _k}} \right)$ represents the inverse cumulative distribution function (CDF) of a Chi-square random variable ${1 - {\varphi _k}}$ with ${2\left( {MB + RN} \right)}$ degrees of freedom. Thus, we complete the transformation of the uncertain region of closed-form. Furthermore,  the channel estimation error can be expressed as
\begin{equation}\label{eq:p42}
\begin{split}
{\bar C16}:& \Delta {\mathbf{x}}_j^H\Delta {{\mathbf{x}}_j} \leqslant \frac{{\psi _k^2\left( {{\mathbf{Tr}}\left( {{{\mathbf{E}}_{d,j}}} \right) + {\mathbf{Tr}}\left( {{{\mathbf{E}}_{f,j}}} \right)} \right)}}{{MB + RN}}, \\
 & \  \forall k \in \mathcal{K}{\text{,}}\forall j \in \mathcal{J}.
\end{split}
\end{equation}

Since (37a), (37b) and ${\bar C16}$ belong to semi-infinite constraints, to obtain the exact equivalent constraints, we first introduce the following lemma [40]:

\emph{Lemma 2 (S-Procedure):} Assume a function
\begin{equation}\label{eq:p43}
{g_i}\left( {\mathbf{x}} \right) \triangleq {{\mathbf{x}}^H}{{\mathbf{C}}_i}{\mathbf{x}} + 2\operatorname{Re} \left\{ {{\mathbf{b}}_i^H{\mathbf{x}}} \right\} + {d_i},i = 1,2,
\end{equation}
where ${{\mathbf{C}}_i} \in {\mathbb{C}^{P \times P}}$, ${{\mathbf{b}}_i} \in {\mathbb{C}^{P \times 1}}$, ${\mathbf{x}} \in {\mathbb{C}^{P \times 1}}$, and ${d_i} \in \mathbb{R}$. Then, the function ${g_1}\left( {\mathbf{x}} \right) \leqslant 0 \Rightarrow {g_2}\left( {\mathbf{x}} \right) \leqslant 0$ holds if and only if there exists
$\kappa  \geqslant 0$ such that
\begin{equation}\label{eq:p44}
\begin{split}
\kappa \left[ {\begin{array}{*{20}{c}}
  {{{\mathbf{C}}_1}}&{{{\mathbf{b}}_i}} \\
  {{\mathbf{b}}_1^H}&{{d_1}}
\end{array}} \right] - \left[ {\begin{array}{*{20}{c}}
  {{{\mathbf{C}}_i}}&{{{\mathbf{b}}_i}} \\
  {{\mathbf{b}}_2^H}&{{d_2}}
\end{array}} \right] \geqslant {\mathbf{0}}.
\end{split}
\end{equation}

For convenience, (37a) can be rewritten as
\begin{equation}\label{eq:p45}
\begin{split}
& \Delta {\mathbf{x}}_j^H{{\mathbf{C}}_{W,k}}\Delta {{\mathbf{x}}_j} + \Delta {\mathbf{x}}_j^H{{\mathbf{C}}_{W,k}}{{{\mathbf{\tilde x}}}_j} + {\mathbf{\tilde x}}_j^H{{\mathbf{C}}_{W,k}}\Delta {{\mathbf{x}}_j} \\
& + {\mathbf{\tilde x}}_j^H{{{\mathbf{C}}_{W,k}}}{{{\mathbf{\tilde x}}}_j} - \varpi _k^j \leqslant 0,\forall k \in \mathcal{K}, j \in \mathcal{J},
\end{split}
\end{equation}
where
\begin{equation}\label{eq:p46}
{{\mathbf{C}}_{W,k}} = \left[ {\begin{array}{*{20}{c}}
  {{{\mathbf{W}}_k}}&{{{\mathbf{W}}_k}{{\mathbf{G}}^H}{\mathbf{\Theta }}} \\
  {{{\mathbf{\Theta }}^H}{\mathbf{G}}{{\mathbf{W}}_k}}&{{{\mathbf{\Theta }}^H}{\mathbf{G}}{{\mathbf{W}}_k}{{\mathbf{G}}^H}{\mathbf{\Theta }}}
\end{array}} \right].
\end{equation}
By Lemma 2, combining (42) and (45), we can obtain an LMI as follows (47), where $\nu  = \frac{1}{{MB + RN}}$. Meanwhile, (36d) can be rewritten as
\setcounter{equation}{47}
\begin{equation}\label{eq:p48}
\begin{split}
 - \Delta {\mathbf{x}}_j^H{{\mathbf{C}}_{L,k}}\Delta {{\mathbf{x}}_j} - \Delta {\mathbf{x}}_j^H{{\mathbf{C}}_{L,k}}{{{\mathbf{\tilde x}}}_j} - {\mathbf{\tilde x}}_j^H{{\mathbf{C}}_{L,k}}\Delta {{\mathbf{x}}_j} \\
 - {\mathbf{\tilde x}}_j^H{{\mathbf{C}}_{L,k}}{{{\mathbf{\tilde x}}}_j} + \chi _k^j - \sigma _{e,j}^2 \leqslant 0,\forall k \in \mathcal{K}, j \in \mathcal{J},
\end{split}
\end{equation}
where
\begin{equation}\label{eq:p49}
\begin{split}
{{\mathbf{C}}_{L,k}} = \left[ {\begin{array}{*{20}{c}}
  {{{\mathbf{L}}_k}}&{{{\mathbf{L}}_k}{{\mathbf{G}}^H}{\mathbf{\Theta }}} \\
  {{{\mathbf{\Theta }}^H}{\mathbf{G}}{{\mathbf{L}}_k}}&{{{\mathbf{\Theta }}^H}{\mathbf{G}}{{\mathbf{L}}_k}{{\mathbf{G}}^H}{\mathbf{\Theta }}}
\end{array}} \right].
\end{split}
\end{equation}
Similar to (47), we can convert (48) to an LMI as (50).

Finally, we can transform the optimization problem ${{\hat {\cal P}}_3}$  into ${{\hat {\cal P}}_4}$  as follows
\setcounter{equation}{50}
\begin{subequations}\label{eq:p51}
\begin{align}
 & {{\hat {\cal P}}_4}  :   \mathop {\max }\limits_{{\mathbf{W}},{\mathbf{V}} ,\boldsymbol{\alpha},\boldsymbol{\beta},{\boldsymbol{\lambda}},{\boldsymbol{\varepsilon}},\boldsymbol{\delta},\boldsymbol{\varsigma},\boldsymbol{\chi},\boldsymbol{\varpi},\boldsymbol{\kappa},\boldsymbol{\omega},{z}  } \ \ z \\
   & {\text{s}}{\text{.t}}{\text{.}}   \ \     C1,    C5,  C6,  \tilde C7,  \bar C8,  C9,  \bar C15,  C17-C20.
\end{align}
\end{subequations}
It can be observed that except for the rank-one constraints of $C5$ and $C6$, the other parts are solvable convex constraints. To this end, we can solve  ${{\hat {\cal P}}_4}$ using SDP technique by removing the rank-one constraint, and Gaussian randomization method can used to obtain the rank-one solution.

\subsection{Fix $({\mathbf{W}},{\mathbf{V}})$ and Solve ${\boldsymbol{\theta }}$}
Based on obtaining ${\mathbf{W}}$ and ${\mathbf{V}}$ at the first section, we rewrite the optimization problem as follows
\begin{subequations}\label{eq:p52}
\begin{align}
 {{\hat {\cal P}}_5} \  &: \mathop {\max }\limits_{ {\bf{\hat Q}} ,\boldsymbol{\hat \alpha} ,\boldsymbol{\hat \beta},{\boldsymbol{\lambda}},{\boldsymbol{\varepsilon}},{z} } \ z \\
  {\text{s}}{\text{.t}}{\text{.}} & \  {\bar C3},{\tilde C4}, {\bar C7} ,{ C13} ,{C14},{\bar C15},{C16}.
\end{align}
\end{subequations}

One can clearly observe that ${{\hat {\cal P}}_5}$ is intractable to solve  due to the coupled variables of ${\bar C15}$ and the semi-infinite constraints of ${\tilde C4}$.  Fortunately, the singular value decomposition (SVD)  method offers a way, and we first deal with ${\bar C15}$. It is obvious that  ${\mathbf{G}}{{\mathbf{D}}_k}{{\mathbf{G}}^H}$ can be written $\sum\nolimits_s {{x_{k,s}}} {{\mathbf{o}}_{k,s,}}{\mathbf{v}}_{k,s}^H$, and thus ${{\mathbf{\Theta }}^H}{\mathbf{G}}{{\mathbf{D}}_k}{{\mathbf{G}}^H}{\mathbf{\Theta }}$ can be expressed as $\sum\nolimits_s {{x_{k,s}}} {\text{diag}}\left( {{{\mathbf{o}}_{k,s}}} \right){\boldsymbol{\theta }} {{\boldsymbol{\theta }}^H}{\text{diag}}\left( {{\mathbf{v}}_{k,s}^H} \right)$, where ${{x_{k,s}}}$, ${{\mathbf{o}}_{k,s}} \in {\mathbb{C}^{RN \times 1}}$ and ${{\mathbf{v}}_{k,s}} \in {\mathbb{C}^{RN \times 1}}$ denote the singular values, left singular vectors and right singular vectors, respectively. Next, we can rewrite it into the following:
\begin{equation}\label{eq:p53}
\sum\nolimits_s {{x_{k,s}}} {\text{diag}}\left( {{{\mathbf{o}}_{k,s}}} \right){\boldsymbol{\theta }} {{\boldsymbol{\theta }} ^H}{\text{diag}}\left( {{\mathbf{v}}_{k,s}^H} \right) = \sum\nolimits_s {{x_{k,s}}} {{\mathbf{O}}_{k,s}}{\mathbf{\hat Q}}{{\mathbf{V}}_{k,s}},
\end{equation}
where ${{\mathbf{O}}_{k,s}} = \left[ {{\text{diag}}\left( {{{\mathbf{o}}_{k,s}}} \right),{\mathbf{0}}} \right]$, and ${{\mathbf{V}}_{k,s}} = {\left[ {{\text{diag}}\left( {{{\mathbf{v}}_{k,s}}} \right),{\mathbf{0}}} \right]^H}$. Furthermore, ${\mathbf{\Theta }}$ can be denoted as ${{\text{diag}}\left( {{{[ {{\mathbf{\hat Q}}} ]}_{1:RN,RN + 1}}} \right)}$, where ${[ {{\mathbf{\hat Q}}} ]_{1:RN,RN + 1}} = {\left[ {{{[ {{\mathbf{\hat Q}}} ]}_{1,RN + 1}}, \ldots ,{{[ {{\mathbf{\hat Q}}} ]}_{RN,RN + 1}}} \right]^T}$.
Therefore, ${{{\mathbf{A}}_k}}$ and ${{{\mathbf{ u}}}_{k,j}}$  in $\bar C15$ can be denoted as (54) and
\setcounter{equation}{54}
\begin{equation}\label{eq:p55}
{{{\mathbf{\hat u}}}_{k,j}} = \left[ {\begin{array}{*{20}{c}}
  {{\sigma _{d,j}}{{\mathbf{D}}_k}{{{\mathbf{\tilde h}}}_j}} \\
  {{\sigma _{f,j}}{\left({{\text{diag}}\left( {{{[ {{\mathbf{\hat Q}}} ]}_{1:RN,RN + 1}}} \right)}\right)^H}{\mathbf{G}}{{\mathbf{D}}_k}{{{\mathbf{\tilde h}}}_j}}
\end{array}} \right],
\end{equation}
respectively. Next, we can convert ${\bar C3}$ into $\tilde C8$ and $\tilde C9$ in (26). Then, we apply Lemma 2 to deal with ${\tilde C4}$ similar to the previous subsection. Here we still use the SVD method to denote ${\mathbf{G}}{{\mathbf{W}}_k}{{\mathbf{G}}^H}$ and ${\mathbf{G}}{{\mathbf{L}}_k}{{\mathbf{G}}^H}$ as $\sum\nolimits_s {{{\hat x}_{k,s}}} {{{\mathbf{\hat o}}}_{k,s}}{\mathbf{\hat v}}_{k,s}^H$ and $\sum\nolimits_s {{{\tilde x}_{k,s}}} {{{\mathbf{\tilde o}}}_{k,s}}{\mathbf{\tilde v}}_{k,s}^H$, respectively, where ${\hat x}_{k,s}$, ${{{\mathbf{\hat o}}}_{k,s}}\in {\mathbb{C}^{RN \times 1}}$, ${\mathbf{\hat v}}_{k,s}\in {\mathbb{C}^{RN \times 1}}$ and ${\tilde x}_{k,s}$, ${{{\mathbf{\tilde o}}}_{k,s}}\in {\mathbb{C}^{RN \times 1}}$, ${\mathbf{\tilde v}}_{k,s}\in {\mathbb{C}^{RN \times 1}}$ represent the corresponding singular values, left and right singular vectors,  respectively. Therefore, we can obtain the following equation:
\begin{subequations}\label{eq:p56}
\begin{align}
 & \sum\nolimits_s {{{\hat x}_{k,s}}} {\text{diag}}\left( {{{{\mathbf{\hat o}}}_{k,s}}} \right){\boldsymbol{\theta }}{{\boldsymbol{\theta }} ^H}{\text{diag}}\left( {{\mathbf{\hat v}}_{k,s}^H} \right) = \sum\nolimits_s {{{\hat x}_{k,s}}} {{{\mathbf{\hat O}}}_{k,s}}{\mathbf{\hat Q}}{{{\mathbf{\hat V}}}_{k,s}},  \\
&  \sum\nolimits_s {{\tilde x_{k,s}}} {\text{diag}}\left( {{{{\mathbf{\tilde o}}}_{k,s}}} \right){\boldsymbol{\theta }} {{\boldsymbol{\theta }} ^H}{\text{diag}}\left( {{\mathbf{\tilde v}}_{k,s}^H} \right) = \sum\nolimits_s {{{\tilde x}_{k,s}}} {{{\mathbf{\tilde O}}}_{k,s}}{\mathbf{\hat Q}}{{{\mathbf{\tilde V}}}_{k,s}},
\end{align}
\end{subequations}
where ${{\mathbf{\hat O}}_{k,s}} = \left[ {{\text{diag}}\left( {{{\mathbf{\hat o}}_{k,s}}} \right),{\mathbf{0}}} \right]$, ${{\mathbf{\hat V}}_{k,s}} = {\left[ {{\text{diag}}\left( {{{\mathbf{\hat v}}_{k,s}}} \right),{\mathbf{0}}} \right]^H}$,
${{\mathbf{\tilde O}}_{k,s}} = \left[ {{\text{diag}}\left( {{{\mathbf{\tilde o}}_{k,s}}} \right),{\mathbf{0}}} \right]$, and ${{\mathbf{\tilde V}}_{k,s}} = {\left[ {{\text{diag}}\left( {{{\mathbf{\tilde v}}_{k,s}}} \right),{\mathbf{0}}} \right]^H}$.
Next, we update (46) and (49) to (57a) and (57b), respectively. Meanwhile, ${{\mathbf{C}}_{W,k}}$ of $C19$ and ${{\mathbf{C}}_{L,k}}$ of $C20$ can be updated by (57a) and (57b), respectively.

\begin{algorithm}[t]
	\renewcommand{\algorithmicrequire}{\textbf{Input:}}
	\renewcommand{\algorithmicensure}{\textbf{Output:}}
    \caption{Proposed Algorithm Based on Imperfect CSI.}
    \label{Algorithm1}
    \begin{algorithmic}[2]
        \REQUIRE ${{\bf{h}}_{b,k}}$, ${{\bf{\tilde h}}_{b,e,j}}$, ${{\bf{f}}_{r,k}}$, ${{\bf{\tilde f}}_{r,e,j}}$, ${{\bf{G}}_{b,r}}$, and ${\sigma _k}$, ${\sigma _e}$.
        \ENSURE Beamforming matrix ${\bf{W}}$, AN matrix ${\bf{V}}$, phase shift matrix ${\bf{\hat Q}}$, and SEE $z$.
        \STATE Initialize ${\bf{W}}^{\left[ t \right]}$, ${\bf{V}}^{\left[ t \right]}$, ${\bf{\hat Q}}^{\left[ t \right]}$,  ${\boldsymbol{\alpha}}^{\left[ t \right]}$, ${\boldsymbol{\beta}}^{\left[ t \right]}$, ${\boldsymbol{\beta}}^{\left[ t-1 \right]}$, ${\boldsymbol{\delta}}^{\left[ t \right]}$, ${\boldsymbol{\varsigma}}^{\left[ t \right]}$, ${z^{\left[ n \right]}}$, $t=0$ and threshold $\tau$;
        \WHILE{ ${z^{\left[ t \right]}}-{z^{\left[ t-1 \right]}}>{\tau}$}
        \STATE $t=t+1$;
        \STATE Update ${\boldsymbol{\rho }}^{\left[ t \right]}$ by (17);
        \STATE Update ${\bf{W}}^{\left[ t \right]}$, ${\bf{V}}^{\left[ t \right]}$  by solving ${{\hat {\cal P}}_4}$;
        \STATE Update $z^{\left[ t \right]}$, ${\bf{\hat Q}}^{\left[ t \right]}$,  by solving ${{\hat {\cal P}}_6}$;
        \ENDWHILE
        \STATE \textbf{return} ${\bf{W}}^{\left[ t \right]}$, ${\bf{\hat Q}}^{\left[ t \right]}$, and $z^{\left[ t \right]}$.
    \end{algorithmic}
\end{algorithm}

Finally, the problem ${{\hat {\cal P}}_5}$ can be rewritten as
\setcounter{equation}{57}
\begin{subequations}\label{eq:p58}
\begin{align}
  {{\hat {\cal P}}_6} : &   \mathop {\max }\limits_{{\bf{\hat Q}} ,\boldsymbol{\hat \alpha} ,\boldsymbol{\hat \beta},{\boldsymbol{\lambda}},{\boldsymbol{\varepsilon}},\boldsymbol{\delta},\boldsymbol{\varsigma},\boldsymbol{\chi},\boldsymbol{\varpi},\boldsymbol{\kappa},\boldsymbol{\omega},{z}  } \ \ z \\
   & {\text{s}}{\text{.t}}{\text{.}}   \  {\bar C7}, {\tilde C8},{\tilde C9},{C13},{C14}, {\bar C15}\{{{{\mathbf{\hat A}}}_k},{{{\mathbf{\hat u}}}_{k,j}}\}, \\
   &  \quad \ \;  {C17},{C18},{C19}\{{{\mathbf{\hat C}}_{W,k}}\},{C20}\{{\mathbf{\hat C}}_{L,k}\}.
\end{align}
\end{subequations}
By using SDP technique to remove the rank-one constraint of $C13$, ${{\hat {\cal P}}_6}$ becomes a convex problem that is easy to solve. Similarly, a feasible solution $\boldsymbol{\theta}^{*}$ can be obtained from ${{{\mathbf{\hat Q}}}^{opt}}$ by using the Gaussian randomization method when it does not satisfy the rank-one constraint. Moreover, we summarize the above procedure as \textbf{Algorithm 2}.

\section{Supplementary Framework}
In this section, we analyze the convergence, optimality and computational complexity of the proposed algorithms.
\subsection{Convergence and Optimality}
For \textbf{Algorithm 1}, we need to alternatively solve ${{\cal P}_4}$ and ${{\cal P}_6}$ until convergence. Since the rank-one constraints are all  dropped, ${{\cal P}_4}$ and ${{\cal P}_6}$ are both convex optimization problems, and thus the KKT solutions can be guaranteed. Similarly, ${{\hat {\cal P}}_4}$ and ${{\hat {\cal P}}_6}$ are solved alternatively for \textbf{Algorithm 2}, and they are also both convex optimization problem after removing the rank-one constraints, and thus the KKT solutions are also guaranteed.   Additionally, since ${\mathbf{W}}$ and ${\mathbf{V}}$ are bounded because of  the limited transmit power and phase shift ${\boldsymbol{\theta }}$ is bounded due to  the unit modulo, ${{\cal P}_0}$ and ${\hat {\cal P}_0}$ both have an upper bound. Based on this, SEE under \textbf{Algorithm 1} and \textbf{Algorithm 2} should be monotonically non-decreasing and converge to a local optimal solution at least, which can be verified in the following simulation results.

\subsection{Computational Complexity Analysis}
In this subsection, we analyze the computational complexity of the proposed algorithms. First, we give an iteration precision $\omega $, and the number of iterations of ${\mathcal{P}_4}$ can be expressed as $\sqrt \Delta  \ln \left( {{1 \mathord{\left/ {\vphantom {1 \omega }} \right. \kern-\nulldelimiterspace} \omega }} \right)$ [41]. There are equivalent $(B+3K+4KJ)$ LMI constraints  and $K$ second-order cone constraints. Therefore, the barrier parameter $\Delta$  can be expressed as $\Delta=B+2MBK+3K+5JK$. The computational complexity of solving ${\mathcal{P}_4}$ is calculated as
\begin{equation}\label{eq:p59}
\mathcal{O}\left( {\sqrt \Delta  \ln \left( {{1 \mathord{\left/
 {\vphantom {1 \omega }} \right.
 \kern-\nulldelimiterspace} \omega }} \right)\left( {{n_0}{n_1} + n_0^2{n_2} + n_0^3} \right)} \right),
\end{equation}
where ${n_0} = \mathcal{O}\left( {2K{M^2}{B^2}} \right)$ and ${2K{M^2}{B^2}}$ represents the number of main optimization variables, ${n_1} =  6K + 2{B^3}{M^3}K + 11JK $, ${n_2} = 2K + 2{M^2}{B^2}K + 7JK $. For ${\mathcal{P}_6}$, there are equivalent $2+K+4KJ+RN$ LMI constraints and $K$ second-order cone constraints. Given an iteration accuracy $\tilde \omega$, the computational complexity of solving ${\mathcal{P}_6}$ is calculated as
\begin{equation}\label{eq:p60}
\mathcal{O}\left( {\sqrt {\tilde \Delta } \ln \left( {{1 \mathord{\left/
 {\vphantom {1 {\tilde \omega }}} \right.
 \kern-\nulldelimiterspace} {\tilde \omega }}} \right)\left( {{{\tilde n}_0}{{\tilde n}_1} + \tilde n_0^2{{\tilde n}_2} + \tilde n_0^3} \right)} \right),
\end{equation}
where the barrier parameter $\tilde \Delta  = 2 + 2RN + 3K + 5KJ$, ${{\tilde n}_0} = \mathcal{O}\left( {{{\left( {RN + 1} \right)}^2}} \right)$, ${{\tilde n}_1} = 5K + 11KJ + \left( {RN + 1} \right) + {\left( {RN + 1} \right)^3}$, and ${{\tilde n}_2} = K + 7KJ + \left( {RN + 1} \right) + {\left( {RN + 1} \right)^2}$.
Therefore, the total computational complexity of \textbf{Algorithm 1} is ${{\cal O}}\left( {m\left( {{n_0}{n_1} + n_0^2{n_2} + n_0^3} \right) + {\tilde m}\left( {{{\tilde n}_0}{{\tilde n}_1} + \tilde n_0^2{{\tilde n}_2} + \tilde n_0^3} \right)} \right)$,
where $m = \sqrt \Delta  \ln \left( {{1 \mathord{\left/ {\vphantom {1 \omega }} \right. \kern-\nulldelimiterspace} \omega }} \right)$
and $\tilde m = \sqrt {\tilde \Delta } \ln \left( {{1 \mathord{\left/  {\vphantom {1 {\tilde \omega }}} \right. \kern-\nulldelimiterspace} {\tilde \omega }}} \right)$.

Similarly, for ${{\hat {\cal P}}_4}$, there are equivalent $B+3K+8KJ$ LMI constraints and $K+KJ$ second-order cone constraints. Given an iteration accuracy $\hat \omega$, the computational complexity of solving ${{\hat {\cal P}}_4}$ is calculated as
\begin{equation}\label{eq:p61}
\mathcal{O}\left( {\sqrt {\hat \Delta } \ln \left( {{1 \mathord{\left/
 {\vphantom {1 {\hat \omega }}} \right.
 \kern-\nulldelimiterspace} {\hat \omega }}} \right)\left( {{{\hat n}_0}{{\hat n}_1} + \hat n_0^2{{\hat n}_2} + \hat n_0^3} \right)} \right),
\end{equation}
where  $\hat \Delta  = B + 3K+ 2MBK  + \left( {10 + 3MB + 3RN} \right)KJ$, ${{\hat n}_0} = \mathcal{O}\left( {2K{M^2}{B^2}} \right)$, ${{\hat n}_1} = 6K + 2{M^3}{B^3}K + 12KJ + {\left( {MB + RN} \right)^3}KJ + 2{\left( {MB + RN + 1} \right)^3}KJ + {\left( {{{\left( {MB + RN} \right)}^2} + \left( {MB + RN} \right)} \right)^2}KJ$,
${{\hat n}_2} = 2K + 8KJ + 2K{M^2}{B^2} + {\left( {MB + RN} \right)^2}KJ + 2{\left( {MB + RN + 1} \right)^2}KJ$.  For ${{\hat {\cal P}}_6}$, there are equivalent $2+RN+K+8KJ$ LMI constraints and $K+KJ$ second-order cone constraints. Given an iteration accuracy $\hat \omega$, the computational  complexity of solving ${{\hat {\cal P}}_6}$ is calculated as
\begin{equation}\label{eq:p62}
\mathcal{O}\left( {\sqrt {\bar \Delta } \ln \left( {{1 \mathord{\left/
 {\vphantom {1 {\bar \omega }}} \right.
 \kern-\nulldelimiterspace} {\bar \omega }}} \right)\left( {{{\bar n}_0}{{\bar n}_1} + \bar n_0^2{{\bar n}_2} + \bar n_0^3} \right)} \right)
\end{equation}
where  $\bar \Delta  = 2 +2RN+ 3K  + \left( {10 + 3MB + 3RN} \right)KJ$, ${{\bar n}_0} = \mathcal{O}\left( {{{\left( {RN + 1} \right)}^2}} \right)$, ${{\bar n}_1} = \left( {RN + 1} \right)+ {\left( {RN + 1} \right)^3} + 5K + 12KJ + {\left( {MB + RN} \right)^3}KJ + 2{\left( {MB + RN + 1} \right)^3}KJ + {\left( {{{\left( {MB + RN} \right)}^2} + \left( {MB + RN} \right)} \right)^2}KJ$, ${{\bar n}_2} = {RN + 1}  + {\left( {RN + 1} \right)^2} + K + 8KJ + 2{\left( {MB + RN + 1} \right)^2}KJ + {\left( {MB + RN} \right)^2}KJ$.
Therefore, the total computational complexity of \textbf{Algorithm 2} is ${{\cal O}}\left( \hat m\left( {{{\hat n}_0}{{\hat n}_1} + \hat n_0^2{{\hat n}_2} + \hat n_0^3} \right) \right. + \left.  \bar m\left( {{{\bar n}_0}{{\bar n}_1} + \bar n_0^2{{\bar n}_2} + \bar n_0^3} \right) \right)$,
where $\hat m = \sqrt {\hat \Delta } \ln \left( {{1 \mathord{\left/ {\vphantom {1 {\hat \omega }}} \right. \kern-\nulldelimiterspace} {\hat \omega }}} \right)$
and $\bar m = \sqrt {\bar \Delta } \ln \left( {{1 \mathord{\left/ {\vphantom {1 {\bar \omega }}} \right. \kern-\nulldelimiterspace} {\bar \omega }}} \right)$.

\section{Simulation Results}
In this section, we provide the simulation results to evaluate the performance of the proposed algorithms. We set $B=2$, $R=2$, $K=2$, and $J=2$. The heights of the  BS,  RIS,  user, and  Eve are 12 m, 8 m, 1.5 m, and 1.5 m, respectively. And their plane coordinates are (0 m, 40($b$-1)+30 m), (65 m, 40($r$-1)+30 m), (60 m, 5($k$-1)+30 m), and (55 m, 5($j$-1)+32 m), respectively. The numbers of each BS antenna and each RIS elements are $M$ = 2 and $N$ = 4, respectively.
We set the maximum transmit power and power amplifier efficiency of each BS to $P_b=15$ dBm and ${\zeta }=1/3$, respectively.
The channel model includes large-scale and small-scale fading [42]. Large scale fading is given by ${L}(d){\rm{ = }}\sqrt {{L_0}{{\left( {\frac{d}{{{d_0}}}} \right)}^{ - \upsilon  }}}, \upsilon  \in \left\{ {{\upsilon _{BU,}},{\upsilon _{BE}},{\upsilon _{BR}},{\upsilon _{BE}}} \right\} $, where $d$, ${L_0}$ and $\upsilon$, respectively, represent the distance between the receiver and the transmitter, the path loss of reference distance ${d_0}=1$ m, and the path loss exponent. Small scale fading model is considered as ${{{\bf{H}}^ * } = \sqrt {\frac{{K'}}{{K' + 1}}} {\bf{H}}_{\text {LoS}}^ *  + \sqrt {\frac{1}{{K' + 1}}} {\bf{H}}_{\text{NLoS}}^ *}, K' \in \left\{ {{K' _{BU,}},{K'_{BE}},{K'_{BR}},{K'_{BE}}} \right\}$, where ${\bf{H}}_{\text{ LoS}}^ *$, ${\bf{H}}_{\text{ NLoS}}^ *$ and $K'$ represent the line-of-sight (LoS) path, the non-LoS path (Rayleigh fading component) and Rayleigh factor respectively. ${\bf{H}}_{\text{ LoS}}^ *$ is expressed as ${\mathbf{H}}_{{\text{LoS}}}^* = {\mathbf{a}}\left( {{\vartheta _{{\text{AoA}}}}} \right){\mathbf{a}}{\left( {{\vartheta _{{\text{AoD}}}}} \right)^H}$, where ${\mathbf{a}}\left( {{\vartheta _{{\text{AoA}}}}} \right) = \exp {\left( {j\frac{{2\pi {d_r}}}{\lambda }\left( {0, \ldots ,\left( {{A_r} - 1} \right)} \right)\sin {\vartheta _{{\text{AoA}}}}} \right)^T}$ and ${\mathbf{a}}\left( {{\vartheta _{{\text{AoD}}}}} \right) = \exp {\left( {j\frac{{2\pi {d_t}}}{\lambda }\left( {0, \ldots ,\left( {{A_t} - 1} \right)} \right)\sin {\vartheta _{{\text{AoD}}}}} \right)^T}$. Here ${A_r}$, ${d_r}$ and ${\vartheta _{{\text{AoA}}}}$, respectively, denote the number of antennas of the receiver, the inter-antenna separation distance and the angle of arrival (AoA), and ${A_t}$, ${d_t}$  and ${\vartheta _{{\text{AoD}}}}$, respectively, denote the number of antennas of the transmitter, the inter-antenna separation distance and the angle of departure (AoD). The maximum outage probability of $k$-th user security rate is ${\varphi _k}=0.1$. We define the maximum normalized error as $\bar \sigma  = \frac{{{{\left\| {{{{\mathbf{\tilde h}}}_{d,e,j}}} \right\|}^2}}}{{{{\left\| {\Delta {{\mathbf{h}}_{d,e,j}}} \right\|}^2}}} = \frac{{{{\left\| {{{{\mathbf{\tilde f}}}_{e,j}}} \right\|}^2}}}{{{{\left\| {\Delta {{{\mathbf{\tilde f}}}_{e,j}}} \right\|}^2}}}$. Other parameters setting can be found in TABLE 1.

\begin{table}[t]
	\caption{List of Key Notations}
	\label{KeyNotations}
	\centering
	\begin{tabular}{m{75pt}|m{140pt}}
		\hline \hline
        Parameters & Values \\ \hline
		Path loss exponent & ${\upsilon _{BU}} = {\upsilon _{BE}} = 3.6$, ${\upsilon _{RU}} = {\upsilon _{RE}} = 2.2$, ${\upsilon _{BR}} = 2.0$   \\ \hline
		Rician channel factor & ${{K'}_{BU}} = {{K'}_{BE}} ={{K'}_{RU}} = {{K'}_{RE}}= 0$, ${{K'}_{BR}} = \infty$ \\ \hline
        The inter-antenna separation distance  & ${d_a}={{\lambda}_a}/2$, where ${{\lambda}_a}$ denotes wavelength \\ \hline
        Path loss at 1 meter & ${L_0} =  - 30$ dB  \\ \hline
        Hardware-dissipated power  & $P_B=100$ mW, $P_U=20$ mW, $P_R=1$ mW \\ \hline
        The noise power  &  ${\sigma _k^2}={\sigma _{e,j}^2}=-80$ dbm \\ \hline
        The maximum normalized error level  &  $\bar \sigma=0.01$ \\ \hline
        The redundancy of the $k$-th user  &  $R_k^{re}=0.5 $ bits/Hz \\ \hline \hline
	\end{tabular}
 \end{table}

For comparison, we first define the following legend:
\begin{itemize}
\item[$\bullet$]
  \emph{Perfect CSI:} Maximizing the minimum user's SEE under perfect CSI (\textbf{Algorithm 1}).
\item[$\bullet$]
  \emph{Imperfect CSI:} Maximizing the minimum user's SEE under perfect CSI (\textbf{Algorithm 2}).
\item[$\bullet$]
  \emph{Perfect CSI without RIS:} Maximizing the minimum user's SEE under perfect CSI without RIS based on \textbf{Algorithm 1}.
\item[$\bullet$]
  \emph{Imperfect CSI without RIS:} Maximizing the minimum user's SEE under imperfect CSI without RIS based on \textbf{Algorithm 2}.
\item[$\bullet$]
  \emph{SSEEM:} Sum SEE maximization (SSEEM) scheme and use the minimum user's SEE as the indicator under perfect CSI.
\item[$\bullet$]
  \emph{Max-min SSE:} maximize the minimum user's SSE scheme and use the minimum user's SEE as the indicator under perfect CSI.
\end{itemize}

%\textbf{Perfect CSI}: to maximize the minimum user's SEE under perfect CSI from Algorithm 1. \textbf{Imperfect CSI}: to maximize the minimum user's SEE under imperfect CSI from Algorithm 2.  \textbf{Perfect CSI without RIS}: there is to execute algorithm 1 without RIS.  \textbf{Imperfect CSI without RIS}: there is to execute algorithm 2 without RIS. \textbf{SSEEM}: there is the sum SEE maximization (SSEEM) scheme with the minimum user SEE as the indicator.  \textbf{Max-min SSE}: there is the maximize the minimum user's SSE scheme with the minimum user SEE as the indicator.

\begin{figure}[t]
  \centering
  \includegraphics[scale = 0.64]{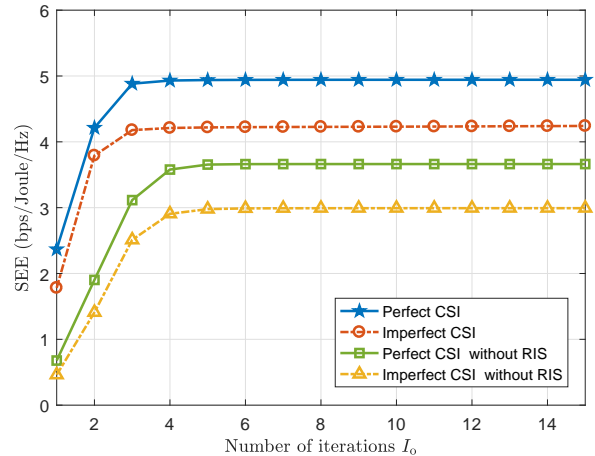}
  \caption{SEE versus the number of iterations.}
  \label{fig.2}
\end{figure}
\emph{1) Relationship between SEE and the number of iterations}: Fig. 2 shows the convergence of different schemes. We can observe that the SEE first increases and then trends to stable after 6 iterations under all schemes. Meanwhile, one can observe that the proposed ``Perfect CSI'' and ``Imperfect CSI'' schemes own higher SEE than the corresponding ``Perfect CSI without RIS'' and ``Imperfect CSI without RIS'' schemes. Additionally,  it is easy to find that the SEE of the ``Imperfect CSI'' scheme is slightly lower than that of the corresponding ``Perfect CSI'' scheme.

\begin{figure}[t]
  \centering
  \includegraphics[scale = 0.64]{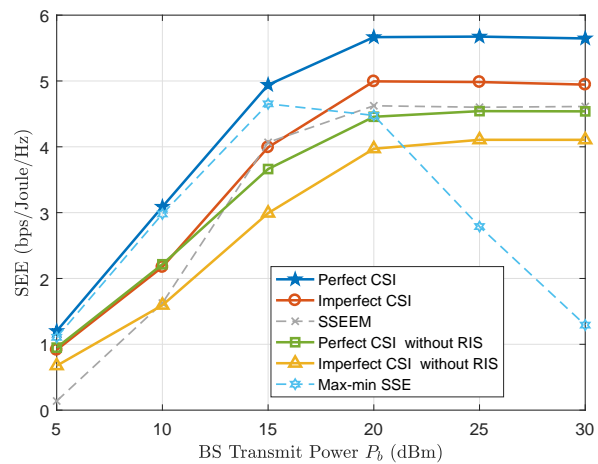}
  \caption{SEE versus BS transmit power $P_b$.}
  \label{fig.3}
\end{figure}

\emph{2) Relationship between SEE and BS transmit power}: We plot Fig. 3  to show the relation between SEE and BS transmit power under different schemes.  One can observe that the SEE first increases and then keeps stable as the transmit power increases under ``Perfect CSI'', ``Imperfect CSI'', ``Perfect CSI without RIS'', ``Imperfect CSI without RIS'' and ``SSEEM'' schemes, while the SEE first increases and then decreases for the ``Max-min SSE'' scheme. This can be explained as follows: when the BS transmit power is low, increasing it can provide a higher SE, thus improving the SEE. However, when the BS transmit power is relative high, such as larger than 20 dBm, the improved SE is very limited when it continues to be increased, and thus the maximum SEE does not increase such as ``Perfect CSI'', ``Imperfect CSI'', ``Perfect CSI without RIS'', ``Imperfect CSI without RIS'' and ``SSEEM'' schemes. However, since the objection of the ``Max-min SSE'' scheme is to maximize the minimum SE, the SEE may decrease as the BS transmit power increases. Meanwhile, we can observe that the SEE under ``Perfect CSI'' scheme owns the highest SEE. In addition, the scheme with RIS is much better in  terms of SEE than the scheme without RIS, which shows  the importance of RISs in improving SEE.

\begin{figure}[t]
  \centering
  \includegraphics[scale = 0.64]{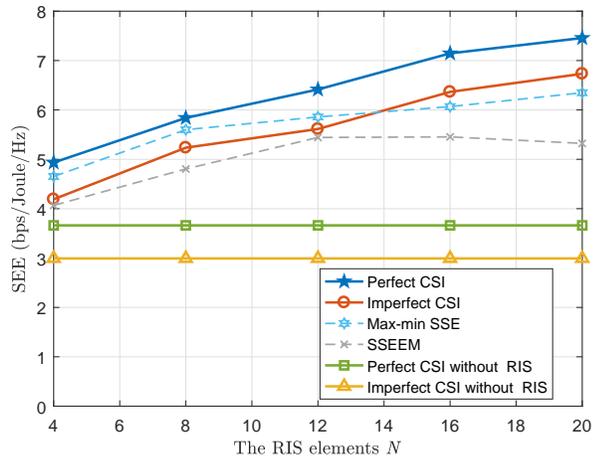}
  \caption{SEE versus RIS elements $N$.}
  \label{fig.4}
\end{figure}
\emph{3) Relationship between SEE and RIS elements $N$}: Fig. 4 shows the SEE versus the number of RIS elements under different schemes.  One can observe that the SEE increases as the number of RIS elements increases under ``Perfect CSI'' and ``Imperfect CSI'' schemes. However, with the increase of RIS elements, since the scheme aims to maximize SE only ensuring the increase of SE,  its SEE is unknown. Fortunately, the SEE of the ``Max-min SSE'' scheme in Fig. 4 can still be improved with the increase of the number of RIS elements. Meanwhile, with the increase of the number of RIS elements, the scheme aims to maximize the sum SEE only ensuring the increase of the sum SEE, and the minimum user's SEE is unknown. For example, the SEE under the ``SSEEM'' scheme in the Fig. 4 increases firstly and then decreases slightly with the increase of the number of RIS elements.  Additionally, the SEE keeps constant for any number of RIS elements under ``Perfect CSI without RIS'' and ``Imperfect CSI without RIS'' schemes, and this is easy to understand.

\begin{figure}[t]
  \centering
  \includegraphics[scale = 0.64]{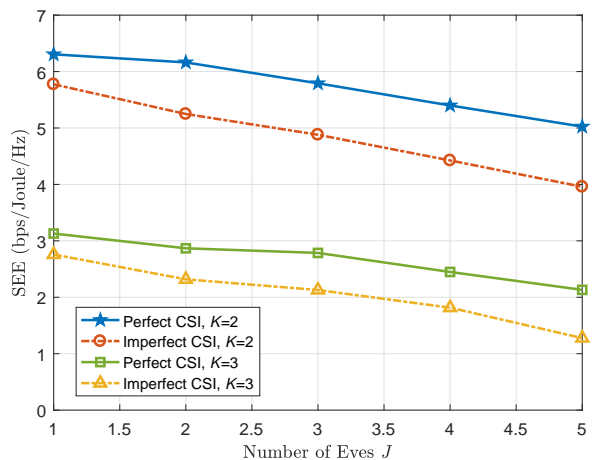}
  \caption{SEE versus the number of Eves $J$.}
  \label{fig.5}
\end{figure}

\emph{4) Relationship between SEE and the number of Eves $J$}:  We redefine the location of the $j$-th Eve and the $k$-th user as (60 m, 8($k$-1)+30 m) and (55 m, 4($j$-1)+31 m), respectively, and illustrate the SEE versus the number of Eves under the proposed schemes based on the different number of users.  We can observe that the SEE decreases as the number of Eves increases under all schemes. This can be explained, as more Eves lead to a higher eavesdropping rate, and thus the SEE decreases accordingly. Meanwhile, it can be found that the SEE decreases with the number of users. This is because more users lead to more serious interference among users, which may change the channel gain of the worst user and decreases the SEE.

\begin{figure}[t]
  \centering
  \includegraphics[scale = 0.64]{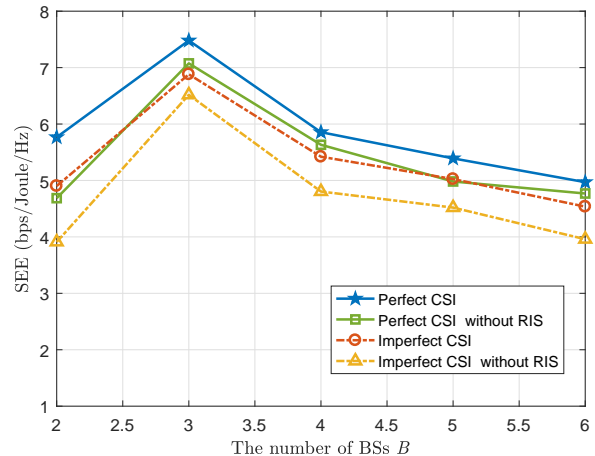}
  \caption{SEE versus the number of BSs $B$.}
  \label{fig.6}
\end{figure}

\begin{figure}[t]
  \centering
  \includegraphics[scale = 0.64]{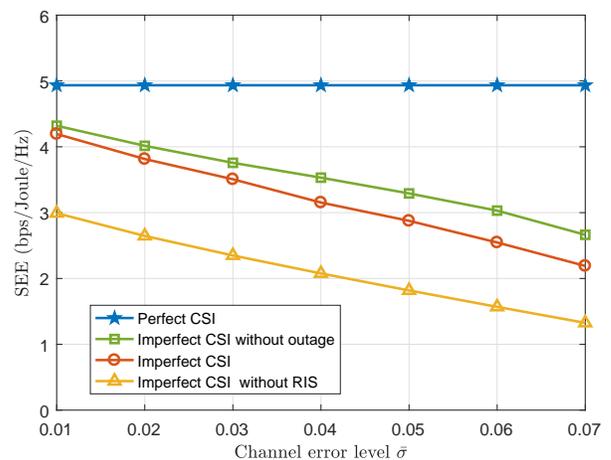}
  \caption{SEE versus the error level $\bar \sigma$.}
  \label{fig.7}
\end{figure}

\emph{5) Relationship between SEE and the number of BSs $B$}:  Here, we reset the position of the $b$-th BS to (0 m, 15$(b-1)$+20 m), and plot Fig. 6 to show the SEE versus the number of BSs $B$. It is obvious that the SEE first increases and then decreases with the number of BSs. This is due to the fact that as the number of BSs grows, more power can be allocated to users and the SEE increases accordingly, but the circuit power consumption also increases. Therefore, when the number of BSs is relatively large, the huge circuit power consumption leads to the decrease of the SEE. In fact, there exists a trade-off between the number of BSs (SSE) and SEE.

\emph{6) Relationship between SEE and the error level $\bar \sigma$}: Fig. 7 shows the SEE versus the error level under different schemes. Here,  we add the ``Imperfect CSI without outage'' scheme  for facilitating comparison. One can observe that the SEE of ``Imperfect CSI'',  ``Imperfect CSI without RIS'' and ``Imperfect CSI without outage'' schemes decreases with the increase of the error level, which is easy to understand. Meanwhile, it is inevitable that the ``Perfect CSI'' scheme remains unchanged with the increase of $\bar \sigma$. Besides, we can observe that the expected performance of the ``Imperfect CSI'' scheme is lower than that of the ``Imperfect CSI without outage'' scheme.

\begin{figure}[t]
  \centering
  \includegraphics[scale = 0.64]{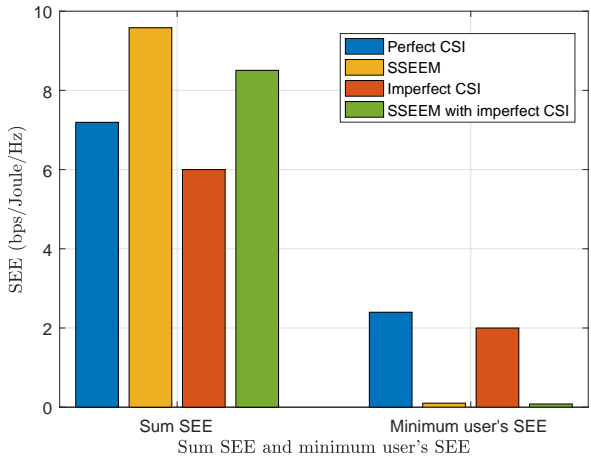}
  \caption{Sum SEE versus minimum user's SEE.}
  \label{fig.8}
\end{figure}

\emph{7) Comparison of SEE fairness}: In Fig. 8, we plot the sum SEE and minimum user's SEE under different schemes. Here, we increase the number of users to three and reset the user locations to (60 m, 70 m), (60 m, 90 m) and (60 m, 120 m), respectively. It can be observed that the sum SEE of ``SSEEM''  and ``SSEEM without imperfect CSI'' schemes are relatively high while the minimum user's SEE is very low. Meanwhile, the sum SEE of our proposed schemes is slightly lower than that of  ``SSEEM''  and ``SSEEM without imperfect CSI'' schemes, but the minimum user's SEE is relatively high. This can be  explained, as the SEEM scheme aims to maximize the sum SEE, but it sacrifices the minimum user's SEE for improving the sum SEE. However, our proposed schemes aim to maximize the minimum user's SEE, and ensure the user's fairness.

\section{Conclusion}
In this paper, we investigated the SEE in the RIS-aided  secure CF networks. We designed a joint active and passive beamforming optimization at BSs and RISs to maximize the minimum  user's SEE, and solved it based on the perfect and imperfect CSI. The simulation results showed that the proposed schemes outperform the existing schemes in term of SEE. Meanwhile, the obtained results also revealed the relations between SEE and the number of BSs, RISs, users, Eves, error level as well as transmit power. It could provide a useful guidance for the application of RIS-aided secure CF networks in future.


\begin{thebibliography}{99}
\bibitem{1}
W. Hao, J. Li, G. Sun, C. Huang, M. Zeng, O. A. Dobre, and C. Yuen. ``Max-min security energy efficiency optimization for RIS-aided cell-free networks,'' \emph{2023 IEEE Int. Conf. Commun. (ICC)}, Under Review.
\bibitem{2}
T. K. Nguyen, H. H. Nguyen, and H. D. Tuan, ``Max-min QoS power control in generalized cell-free massive MIMO-NOMA with optimal backhaul combining,''  \emph{IEEE Trans. Veh. Technol.}, vol. 69, no. 10, pp. 10949-10964, Oct. 2020.
\bibitem{3}
Z. Zhang and L. Dai, ``A joint precoding framework for wideband reconfigurable intelligent surface-aided cell-free network,'' \emph{IEEE Trans. Signal Process.}, vol. 69, pp. 4085-4101, Jun. 2021.
\bibitem{4}
H. Q. Ngo \emph{et al.},  ``Cell-free massive MIMO versus small cells,'' \emph{IEEE Trans. Wireless Commun.}, vol. 16, no. 3, pp. 1834-1850, Mar. 2017.
\bibitem{5}
Y. Fang, L. Qiu, X. Liang, and C. Ren, ``Cell-free massive MIMO systems with oscillator phase noise: Performance analysis and power control,'' \emph{IEEE Trans. Veh. Technol.}, vol. 70, no. 10, pp. 10048-10064, Oct. 2021.
\bibitem{6}
T. K. Nguyen, H. H. Nguyen, and H. D. Tuan, ``Max-min QoS power control in generalized cell-free massive MIMO-NOMA with optimal backhaul combining,''  \emph{IEEE Trans. Veh. Technol.}, vol. 69, no. 10, pp. 10949-10964, Oct. 2020.
\bibitem{7}
Q. N. Le, V. -D. Nguyen, O. A. Dobre, and R. Zhao, ``Energy efficiency maximization in RIS-aided cell-free network with limited backhaul,'' \emph{IEEE Commun. Lett.}, vol. 25, no. 6, pp. 1974-1978, Jun. 2021.
\bibitem{8}
W. Tang \emph{et al.}, ``Wireless communications with reconfigurable intelligent surface: Path loss modeling and experimental measurement,'' \emph{IEEE Trans. Wireless Commun.}, vol. 20, no. 1, pp. 421-439, Jan. 2021.
\bibitem{9}
L. Dong and H. Wang, ``Secure MIMO transmission via intelligent reflecting surface,'' \emph{IEEE Wireless Commun. Lett.}, vol. 9, no. 6, pp. 787-790, Jun. 2020.
\bibitem{10}
Y. Han, X. Li, W. Tang, S. Jin, Q. Cheng, and T. J. Cui, ``Dual-polarized RIS-assisted mobile communications,'' \emph{IEEE Trans. Wireless Commun.}, vol. 21, no. 1, pp. 591-606, Jan. 2022.
\bibitem{11}
L. Zhai, Y. Zou, J. Zhu, and B. Li, ``Improving physical layer security in IRS-aided WPCN multicast systems via stackelberg game,'' \emph{IEEE Trans. Commun.}, vol. 70, no. 3, pp. 1957-1970, Mar. 2022.
\bibitem{12}
H. Yu, S. Guo, Y. Yang, L. Ji, and Y. Yang, ``Secrecy energy efficiency optimization for downlink two-user OFDMA networks with SWIPT,'' \emph{IEEE Syst. J.}, vol. 13, no. 1, pp. 324-335, Mar. 2019. %IEEE Systems Journal
\bibitem{13}
S. Yun, J.-M. Kang, I.-M. Kim, and J. Ha, ``Deep artificial noise: Deep learning-based precoding optimization for artificial noise scheme,'' \emph{IEEE Trans. Veh. Technol.}, vol. 69, no. 3, pp. 3465-3469, Mar. 2020.
\bibitem{14}
A. Khisti and D. Zhang, ``Artificial-noise alignment for secure multicast using multiple antennas,'' \emph{IEEE Commun. Lett.}, vol. 17, no. 8, pp. 1568-1571, Aug. 2013.
\bibitem{15}
Y. Gu, Z. Wu, Z. Yin, and X. Zhang, ``The secrecy capacity optimization artificial noise: A new type of artificial noise for secure communication in MIMO system,''  \emph{IEEE Access}, vol. 7, pp. 58353-58360, Mar. 2019.
\bibitem{16}
W. Hao, J. Li, G. Sun, M. Zeng, and O. A. Dobre, ``Securing reconfigurable intelligent surface-aided cell-free networks,''  \emph{IEEE Trans. Inf. Forensic Secur.}, vol. 17, pp. 3720-3733, Oct. 2022.
\bibitem{17}
S. Elhoushy, M. Ibrahim, and W. Hamouda, ``Exploiting RIS for limiting information leakage to active eavesdropper in cell-free massive MIMO,''  \emph{IEEE Wireless Commun. Lett.}, vol. 11, no. 3, pp. 443-447, Mar. 2022.
\bibitem{18}
Z. Zhang, C. Zhang, C. Jiang, F. Jia, J. Ge, and F. Gong, ``Improving physical layer security for reconfigurable intelligent surface aided NOMA 6G networks,'' \emph{IEEE Trans. Veh. Technol.}, vol. 70, no. 5, pp. 4451-4463, May 2021.
\bibitem{19}
X. Xia \emph{et al.}, ``Joint uplink power control, downlink beamforming, and mode selection for secrecy cell-free massive MIMO with network-assisted full duplexing,'' \emph{IEEE Syst. J.}, to appear, Jul. 2022.
\bibitem{20}
X. Zhang, D. Guo, K. An, and B. Zhang, ``Secure communications over cell-free massive MIMO networks with hardware impairments,'' \emph{IEEE Syst. J.}, vol. 14, no. 2, pp. 1909-1920, Jun. 2020.
\bibitem{21}
M. Alageli \emph{et al.},  ``Optimal downlink transmission for cell-free SWIPT massive MIMO systems with active eavesdropping,'' \emph{IEEE Trans. Inf. Forensic Secur.}, vol. 15, pp. 1983-1998, Nov. 2020.
\bibitem{22}
Y. Zhang, W. Xia, G. Zheng, H. Zhao, L. Yang, and H. Zhu, ``Secure transmission in cell-free massive MIMO With low-resolution DACs over rician fading channels,''  \emph{IEEE Trans. Commun.}, vol. 70, no. 4, pp. 2606-2621, Apr. 2022.
\bibitem{23}
Y. Sun \emph{et al.}, ``Energy-efficient hybrid beamforming for multilayer RIS-assisted secure integrated terrestrial-aerial networks,''  \emph{IEEE Trans. Commun.}, vol. 70, no. 6, pp. 4189-4210, Jun. 2022.
\bibitem{24}
L. Dong, H.-M. Wang, and J. Bai, ``Active reconfigurable intelligent surface aided secure transmission,'' \emph{IEEE Trans. Veh. Technol.}, vol. 71, no. 2, pp. 2181-2186, Feb. 2022.
\bibitem{25}
J. Zhang, H. Du, Q. Sun, B. Ai, and D. W. K. Ng, ``Physical layer security enhancement with reconfigurable intelligent surface-aided networks,''  \emph{IEEE Trans. Inf. Forensic Secur.}, vol. 16, pp. 3480-3495, May 2021.
\bibitem{26}
Y. Han, N. Li, Y. Liu, T. Zhang, and X. Tao, ``Artificial noise aided secure NOMA communications in STAR-RIS networks,''  \emph{IEEE Wirel. Commun. Lett.}, vol. 11, no. 6, pp. 1191-1195, Jun. 2022.
\bibitem{27}
J. Li, S. Xu, J. Liu, Y. Cao, and W. Gao, ``Reconfigurable intelligent surface enhanced secure aerial-ground communication,'' \emph{IEEE Trans. Commun.}, vol. 69, no. 9, pp. 6185-6197, Sep. 2021.
%\bibitem{22}
%U. Siddique, H. Tabassum, and E. Hossain, ``Downlink spectrum allocation for in-band and out-band wireless backhauling of full-duplex small cells,'' \emph{IEEE Trans. Commun.}, vol. 65, no. 8, pp. 3538-3554, Aug. 2017.
\bibitem{28}
J. Li, L. Zhang, K. Xue, Y. Fang, and Q. Sun, ``Secure transmission by leveraging multiple intelligent reflecting surfaces in MISO systems,'' \emph{IEEE. Trans. Mob. Comput.}, to appear, Sep. 2021.
\bibitem{29}
H. Niu, Z. Chu, F. Zhou, Z. Zhu, M. Zhang, and K.-K. Wong, ``Weighted sum secrecy rate maximization using intelligent reflecting surface,"  \emph{IEEE Trans. Commun.}, vol. 69, no. 9, pp. 6170-6184, Sept. 2021.
\bibitem{30}
Y . Han, S. Zhang, L. Duan, and R. Zhang, ``Cooperative double-IRS aided communication: Beamforming design and power scaling,'' \emph{IEEE Wireless Commun. Lett.}, vol. 9, no. 8, pp. 1206-1210, Aug. 2020.
\bibitem{31}
C. Pan \emph{et al.}, ``Intelligent reflecting surface aided MIMO broadcasting for simultaneous wireless information and power transfer,"  \emph{IEEE J. Sel. Area. Comm.}, vol. 38, no. 8, pp. 1719-1734, Aug. 2020.
\bibitem{32}
K. Shen and W. Yu, ``Fractional programming for communication systems---Part I: Power control and beamforming,'' \emph{IEEE Trans. Signal Process.}, vol. 66, no. 10, pp. 2616-2630, May, 2018.
\bibitem{33}
P. Song, G. Scutari, F. Facchinei, and L. Lampariello, ``D3M: Distributed multi-cell multigroup multicasting,'' in \emph{Proc. IEEE Int. Conf. Acoustics, Speech and Signal Processing (ICASSP)}, Shanghai, 2016, pp. 3741-3745.
\bibitem{34}
Y. Xu, H. Xie, Q. Wu, C. Huang, and C. Yuen, ``Robust max-min energy efficiency for RIS-aided HetNets with distortion noises,''  \emph{IEEE Trans. Commun.}, vol. 70, no. 2, pp. 1457-1471, Feb. 2022.
\bibitem{35}
J. Huang and A. L. Swindlehurst, ``Robust secure transmission in MISO channels based on worst-case optimization,'' \emph{IEEE Trans. Signal Process.}, vol. 60, no. 4, pp. 1696-1707, Apr. 2012.
\bibitem{36}
Q. Li and L. Yang, ``Artificial noise aided secure precoding for MIMO untrusted two-way relay systems with perfect and imperfect channel state information,''  \emph{IEEE Trans. Inf. Forensic Secur.}, vol. 13, no. 10, pp. 2628-2638, Oct. 2018.
\bibitem{37}
Z. Li, S. Wang, M. Wen, and Y.-C. Wu, ``Secure multicast energy-efficiency maximization with massive RISs and uncertain CSI: First-order algorithms and convergence analysis,'' \emph{IEEE Trans. Wireless Commun.}, vol. 21, no. 9, pp. 6818-6833, Sep. 2022.
\bibitem{38}
G. Zhou, C. Pan, H. Ren, K. Wang, and A. Nallanathan, ``A framework of robust transmission design for IRS-aided MISO communications with imperfect cascaded channels,'' \emph{IEEE Trans. Signal Process.}, vol. 68, pp. 5092-5106, Aug. 2020.
\bibitem{39}
Q. Li and W.-K. Ma, ``Spatially selective artificial-noise aided transmit optimization for MISO multi-eves secrecy rate maximization,'' \emph{IEEE Trans. Signal Process.}, vol. 61, no. 10, pp. 2704-2717, May 2013.
\bibitem{40}
S. P. Boyd and L. V andenberghe, \emph{Convex Optimization}. Cambridge, U.K.: Cambridge Univ. Press, 2004.
\bibitem{41}
K.-Y . Wang, A.-C. So, T.-H. Chang, W.-K. Ma, and C.-Y. Chi, ``Outage constrained robust transmit optimization for multiuser MISO downlinks: Tractable approximations by conic optimization,''  \emph{IEEE Trans. Signal Process.}, vol. 62, no. 21, pp. 5690-5705, Nov. 2014.
\bibitem{42}
C. Pan \emph{et al.}, ``Intelligent reflecting surface aided MIMO broadcasting for simultaneous wireless information and power transfer,'' \emph{IEEE J. Sel. Areas Commun.}, vol. 38, no. 8, pp. 1719-1734, Aug. 2020.



\end{thebibliography}
\end{document}